\documentclass[11pt,preprintnumbers,superscriptaddress,amsmath,amssymb,nofootinbib]{revtex4}

\usepackage{graphicx}
\usepackage{dcolumn}
\usepackage{bm}
\usepackage{amssymb}
\usepackage{amsmath}
\usepackage{epsfig}    
\usepackage{color}
\usepackage{slashed}

\begin{document} 

\title{Testing the custodial symmetry in the Higgs sector of the Georgi-Machacek model}
\author{Cheng-Wei Chiang}
\email{chengwei@ncu.edu.tw}
\affiliation{Department of Physics and Center for Mathematics and Theoretical Physics,
National Central University, Chungli, Taiwan 32001, ROC}
\affiliation{Institute of Physics, Academia Sinica, Taipei, Taiwan 11529, ROC}
\affiliation{Physics Division, National Center for Theoretical Sciences, Hsinchu, Taiwan 30013, ROC}
\author{Kei Yagyu}
\email{keiyagyu@ncu.edu.tw}
\affiliation{Department of Physics and Center for Mathematics and Theoretical Physics,
National Central University, Chungli, Taiwan 32001, ROC}

\begin{abstract}
We study how the custodial symmetry in the Higgs sector of the Georgi-Machacek (GM) model can be tested at the LHC.  As the minimal extension of the Higgs triplet model, in which tiny neutrino masses are generated via the Type-II Seesaw Mechanism, the GM model keeps 
the electroweak $\rho$ parameter at unity at tree level.  In the GM model, there are 5-plet ($H_5$), 3-plet ($H_3$) and singlet ($H_1$) Higgs bosons under the classification of the custodial $SU(2)_V$ symmetry, in addition to the standard model-like Higgs boson ($h$).
These new Higgs bosons have the following characteristic features at the tree level: (1) the masses of the Higgs bosons belonging to the same $SU(2)_V$ multiplet are degenerate; and (2) $H_5$ and $H_1$ couple to the electroweak gauge bosons but not SM quarks, whereas $H_3$ couples to the quarks but not the gauge bosons.  We find that the $H_5$ production from the weak vector boson fusion process and the Drell-Yan process associated with $H_3$ are useful in testing the custodial symmetry of the Higgs sector at the LHC.  In addition, these processes can also be used to discriminate from other models that contain singly-charged Higgs bosons and extra neutral Higgs bosons.  We also investigate a possible enhancement in the $h\to \gamma\gamma$ as well as $h\to Z\gamma$ decays. 
\end{abstract}

\maketitle

\section{introduction \label{sec:intro}}

Recently, a new particle of mass about 125 GeV has been discovered at the CERN Large Hadron Collider (LHC) with a total production rate consistent with that of the standard model (SM) Higgs boson~\cite{Higgs_CMS, Higgs_ATLAS}.  Confirming this Higgs-like particle as the one responsible for the electroweak symmetry breaking is of paramount importance in particle physics because, for one thing, it explains the origin of mass for elementary particles.  While further detailed examinations are required, the current LHC data show some deviations in the pattern of its decay branching ratios from the SM expectation.  This leads to the speculation that the Higgs sector may not as simple as the one in SM.

In certain new physics models such as supersymmetry, 
the Higgs sector has to be extended with additional nontrivial isospin $SU(2)_L$ scalar multiplets for consistency or to explain new phenomena.  
Such an extension also holds the capacity to provide additional CP-violating sources for low-energy phenomena as well as baryon asymmetry of the Universe.  For example, the two-Higgs doublet model (2HDM) \cite{2HDM} is an extensively studied prototype in which an additional scalar doublet is introduced.  
$SU(2)_L$ triplet Higgs fields also occur in some new physics models, such as the left-right symmetric model~\cite{LRmodels} and little Higgs models~\cite{little-Higgs}. 
By introducing a complex triplet Higgs field, it is possible to have an effective dimension-5 operator for generating tiny Majorana mass for neutrinos.  Therefore, it is important to determine the true Higgs sector in order to exactly know what kind of new physics models exist at the TeV or higher energy scales.
In this paper, we want to focus on the phenomenology of the extended Higgs sector in the
model proposed by Georgi and Machacek (GM) \cite{GM} in mid-80s.  We investigate how one can distinguish it from the other Higgs-extended models at the LHC.

The GM model contains a Higgs doublet field $\Phi$ and a triplet field $\Delta$, with the latter containing a hypercharge $Y=1$ component and a $Y=0$ component.  The model is of great interest because it can provide tiny mass to neutrinos {\`a} la the Seesaw Mechanism, dubbed the Type-II Seesaw~\cite{typeII}.  Moreover, it has been shown that the Higgs potential in this model can be constructed to maintain a custodial $SU(2)_V$ symmetry at the tree level \cite{Chanowitz}, keeping the electroweak $\rho$ parameter at unity to be consistent with the experimental constraint.  
In the model, there are 5-plet Higgs bosons $H_5$ (=$H_5^{\pm\pm},H_5^\pm, H_5^0$), 3-plet Higgs bosons $H_3$ (=$H_3^{\pm},H_3^0$) and singlet Higgs boson $H_1^0$ under the classification of the $SU(2)_V$ symmetry. 
The masses of the Higgs bosons belonging to the same $SU(2)_V$ multiplet are the same at the tree level as the consequence of custodial symmetry.

The doubly-charged Higgs boson $H_5^{\pm\pm}$, for example, is an important but not unique feature of the model.  Finding particles in one Higgs multiplet and checking their (near) mass degeneracy would better verify the model.  Strategies of discovering such Higgs bosons, however, depend largely on the vacuum expectation value (VEV) of the Higgs triplet field, $v_\Delta$.

In the minimal Higgs triplet model (HTM) where only one additional complex Higgs triplet is introduced, the doubly-charged Higgs bosons couples dominantly to a pair of like-sign leptons when $v_\Delta \alt 10^{-4}$ GeV.  The collider phenomenology of this scenario has been extensively studied recently \cite{HTMll}.  The doubly-charged Higgs boson has been searched for at the Tevatron \cite{Tevatronll} and the LHC \cite{LHCll} by looking for like-sign lepton pairs with the same or different flavors.  A lower mass bound of about $400$ GeV has been obtained for most scenarios.  On the other hand, the doubly-charged Higgs bosons couples dominantly to a pair of like-sign $W$ bosons when $v_\Delta \agt 10^{-4}$ GeV
\footnote{
When there is a non-zero mass splitting among the scalar bosons in the triplet Higgs field and the doubly-charged Higgs boson mass is the heaviest, the cascade decays of the doubly-charged Higgs boson become dominant.  Phenomenology of this scenario has been discussed in Refs.~\cite{H++-cascade}.}.  
This possibility is less explored experimentally.  Besides, the triplet VEV in the HTM is constrained by the $\rho$ parameter to be less than a few GeV, limiting significantly the discovery reach at the LHC.

In the GM model, a larger triplet VEV is allowed due to the custodial symmetry.  
It is therefore interesting to consider signatures of the like-sign gauge boson decays.  
In Ref.~\cite{GVW1}, collider phenomenology of the GM model has been discussed in the case of light triplet-like Higgs bosons, {\it e.g.}, less than 100 GeV.
A recent study by one of the authors and collaborators \cite{Chiang_Nomura_Tsumura} finds that with $v_\Delta = 55$ GeV and appropriate cuts, the current LHC can reach up to $450$ GeV for the doubly-charged Higgs mass.  
In this work, we further explore consequences of the custodial symmetry in the Higgs sector of the GM model and study the phenomenology of its entire Higgs sector at the LHC.  
We find that the single production of $H_5$ via the weak vector boson fusion process is useful 
to test the mass degeneracy among the $H_5$ bosons.  We also find that the Drell-Yan process, where $H_5$ and $H_3$ are 
simultaneously produced can be used to check the mass degeneracy among $H_3$.

The structure of this paper is organized as follows.  We review the GM model in Section~\ref{sec:model}.  The Higgs bosons are first classified according to their group representations under the custodial symmetry.  We then consider possible mixings between the two triplets and between the two singlets, and work out their masses.  A mass relation among the Higgs bosons of different representations is obtained in the decoupling limit when the triplet VEV vanishes.  Finally, we show the Yukawa couplings between SM fermions and the physical Higgs bosons.  In Section~\ref{sec:constraints}, we consider both theoretical constraints of perturbative unitarity and vacuum stability and the experimental constraint from the $Z$-pole data of $Z \to b \bar b$ decay at one-loop level.  In particular, they impose bounds on the triplet VEV and the Higgs triplet mass.  In Section~\ref{sec:decays}, we discuss in detail how the Higgs bosons decay in scenarios with or without hierarchy in the masses of the physical Higgs singlet, 3-plet, and 5-plet.  The collider phenomenology of the model can be drastically different in different regions of the $v_\Delta$-$\Delta m$ ($\Delta m$ is the mass difference 
between $H_5$ and $H_3$) space.  
Section~\ref{sec:pheno} discusses how the Higgs bosons can be searched for at the LHC.  Finally, we compute the decay rates of $h \to \gamma\gamma$ and $Z\gamma$ in the model in Section~\ref{sec:diphoton}.  Our findings are summarized in Section~\ref{sec:summary}.

\section{The Model \label{sec:model}}

In the GM model, the Higgs sector is composed of the SM isospin doublet Higgs field $\phi$ with hypercharge $Y=1/2$ and two isospin triplet Higgs fields $\chi$ with $Y=1$ and $\xi$ with $Y=0$. 
%
These fields can be expressed in the form:
\begin{align}
\Phi=\left(
\begin{array}{cc}
\phi^{0*} & \phi^+ \\
\phi^- & \phi^0
\end{array}\right),\quad 
\Delta=\left(
\begin{array}{ccc}
\chi^{0*} & \xi^+ & \chi^{++} \\
\chi^- & \xi^0 & \chi^{+} \\
\chi^{--} & \xi^- & \chi^{0} 
\end{array}\right), \label{eq:Higgs_matrices}
\end{align}
where $\Phi$ and $\Delta$ are transformed under $SU(2)_L\times SU(2)_R$ as
$\Phi\to U_L\Phi U_R^\dagger$ and $\Delta\to U_L\Delta U_R^\dagger$ with
$U_{L,R}=\exp(i\theta_{L,R}^aT^a)$ and $T^a$ being the $SU(2)$ generators. 
The neutral components in Eq.~(\ref{eq:Higgs_matrices}) can be parametrized as 
\begin{align}
\phi^0=\frac{1}{\sqrt{2}}(\phi_r+v_\phi+i\phi_i), \quad 
\chi^0=\frac{1}{\sqrt{2}}(\chi_r+i\chi_i)+v_\chi,\quad
\xi^0=\xi_r+v_\xi, \label{eq:neutral}
\end{align}
where $v_\phi$, $v_\chi$ and $v_\xi$ are the VEV's for $\phi^0$, 
$\chi^0$ and $\xi^0$, respectively. 
When the two triplet VEV's $v_\chi$ and $v_\xi$ are taken to be the same, {\it i.e.}, $v_\chi=v_\xi \equiv v_\Delta$, 
the $SU(2)_L\times SU(2)_R$ symmetry is reduced to the custodial $SU(2)_V$ symmetry. 
The phase convention for the component scalar fields are chosen 
to be $\chi^{--}=(\chi^{++})^*$, $\phi^-=-(\phi^+)^*$, $\chi^-=-(\chi^+)^*$, $\xi^-=-(\xi^+)^*$ and $\xi^0=(\xi^0)^*$. 

The relevant Lagrangian involving the Higgs fields can be written as
\begin{align}
\mathcal{L}_{\text{GM}}
=\mathcal{L}_{\text{kin}}+\mathcal{L}_{Y}+\mathcal{L}_{\nu}-V_H, 
\end{align}
where $\mathcal{L}_{\text{kin}}$, $\mathcal{L}_{Y}$, $\mathcal{L}_{\nu}$ and $V_H$ 
are the kinetic term, the Yukawa interaction between 
$\phi$ and the fermions, the neutrino Yukawa interaction between $\chi$ and the lepton doublets, 
and the Higgs potential, respectively. 

The most general Higgs potential invariant
under the $SU(2)_L\times SU(2)_R\times U(1)_Y$ symmetry in terms of the fields defined in Eq.~(\ref{eq:Higgs_matrices}) is
\begin{align}
V_H&=m_1^2\text{tr}(\Phi^\dagger\Phi)+m_2^2\text{tr}(\Delta^\dagger\Delta)
+\lambda_1\text{tr}(\Phi^\dagger\Phi)^2
+\lambda_2[\text{tr}(\Delta^\dagger\Delta)]^2
+\lambda_3\text{tr}[(\Delta^\dagger\Delta)^2]\notag\\
&+\lambda_4\text{tr}(\Phi^\dagger\Phi)\text{tr}(\Delta^\dagger\Delta)
+\lambda_5\text{tr}\left(\Phi^\dagger\frac{\tau^a}{2}\Phi\frac{\tau^b}{2}\right)
\text{tr}(\Delta^\dagger t^a\Delta t^b)\notag\\
&+\mu_1\text{tr}\left(\Phi^\dagger \frac{\tau^a}{2}\Phi\frac{\tau^b}{2}\right)(P^\dagger \Delta P)^{ab}
+\mu_2\text{tr}\left(\Delta^\dagger t^a\Delta t^b\right)(P^\dagger \Delta P)^{ab}, \label{eq:pot}
\end{align}
where $\tau^a$ are the Pauli matrices, $t^a$ are the $3\times 3$ matrix representation of the $SU(2)$ generators given by 
\begin{align}
t_1=\frac{1}{\sqrt{2}}\left(
\begin{array}{ccc}
0 & 1 & 0 \\
1 & 0 & 1 \\
0 & 1 & 0
\end{array}\right),\quad t_2=\frac{1}{\sqrt{2}}\left(
\begin{array}{ccc}
0 & -i & 0 \\
i & 0 & -i \\
0 & i & 0
\end{array}\right),\quad
t_3=\left(
\begin{array}{ccc}
1 & 0 & 0 \\
0 & 0 & 0 \\
0 & 0 & -1
\end{array}\right),
\end{align}
and the matrix $P$ is defined as 
\begin{align}
P=\left(
\begin{array}{ccc}
-1/\sqrt{2} & i/\sqrt{2} & 0 \\
0 & 0 & 1 \\
1/\sqrt{2} & i/\sqrt{2} & 0
\end{array}\right). 
\end{align}
As in the HTM, 
the SM electroweak symmetry breaking can induce the triplet field to develop a VEV $v_\Delta$ through the $\mu_1$ term in the Higgs potential. 
To our knowledge, most of the previous analyses ignore both $\mu_1$ and $\mu_2$ interactions in their phenomenology studies.  We will keep these terms in this work.

Using the tadpole conditions, 
\begin{align}
\left.\frac{\partial V_H }{\partial \phi_r}\right|_0=0,\quad 
\left.\frac{\partial V_H }{\partial \xi_r}\right|_0=0,\quad 
\left.\frac{\partial V_H }{\partial \chi_r}\right|_0=0, \label{eq:vc}
\end{align}
the parameters $m_1^2$ and $m_2^2$ can be eliminated as 
\begin{subequations}
\label{eq:m1m2}
\begin{align}
m_1^2 & = -v^2\left(2c_H^2\lambda_1+\frac{3}{8}s_H^2\lambda_4+\frac{3}{16}s_H^2\lambda_5\right)+\frac{3}{8}s_H^2M_1^2,\\
m_2^2 & = -v^2\left(\frac{3}{4}s_H^2\lambda_2+\frac{1}{4}s_H^2\lambda_3+c_H^2\lambda_4+\frac{1}{2}c_H^2\lambda_5\right)
+\frac{1}{2}c_H^2M_1^2+\frac{1}{4}M_2^2, \label{eq:m2}
\end{align}
\end{subequations}
where $v^2=v_\phi^2+8v_\Delta^2=1/(\sqrt{2}G_F)$ and $\tan\theta_H=2\sqrt{2}v_\Delta/v_\phi$
with $s_H=\sin\theta_H$ and $c_H=\cos\theta_H$. 
In Eq.~(\ref{eq:m1m2}), we introduce $M_1^2$ and $M_2^2$ 
as
\begin{align}
M_1^2=-\frac{v}{\sqrt{2}s_H}\mu_1,\quad 
M_2^2=-3\sqrt{2}s_Hv\mu_2. 
\end{align}
The second and third conditions in Eq.~(\ref{eq:vc}) give 
the same constraint in Eq.~(\ref{eq:m2}) as long as $v_\chi = v_\xi$. 

Before we discuss the mass matrices and the mass eigenstates for the Higgs bosons, 
it is convenient to classify the Higgs boson states according to the custodial $SU(2)_V$ symmetry. 
The triplet field $\Delta$, which can be understood as a 
$\bm{3}\otimes \bm{3}$ representation of the $SU(2)_V$ multiplet, 
can be decomposed into the irreducible representations  
$\bm{5}\oplus \bm{3}\oplus \bm{1}$. 
Likewise, the doublet field $\Phi$ being the 
$\bm{2}\otimes \bm{2}$ representation of the $SU(2)_V$ multiplet, 
can be decomposed into $\bm{3}\oplus  \bm{1}$. 
The $\bm{3}$ representation of $\Phi$ can be identified as the Nambu-Goldstone (NG) bosons of the SM
as long as there is no mixing 
between the $\bm{3}$ representations of $\Delta$ and $\Phi$. 
The 5-plet ($H_5^{\pm\pm}$, $H_5^\pm$ and $H_5^0$), the 3-plet ($\tilde{H}_3^\pm$ and $\tilde{H}_3^0$) 
and the singlet ($\tilde{H}_1^0$) originating from $\Delta$ can be related to the original component fields as 
\begin{align}
&H_5^{\pm\pm} = \chi^{\pm\pm},\quad H_5^\pm = \frac{1}{\sqrt{2}}(\chi^\pm-\xi^\pm),\quad H_5^0=\frac{1}{\sqrt{3}}(\chi_r-\sqrt{2}\xi_r), \notag\\
&\tilde{H}_3^\pm = \frac{1}{\sqrt{2}}(\chi^\pm+\xi^\pm),\quad \tilde{H}_3^0 = \chi_i,\notag\\
&\tilde{H}_1^{0} = \frac{1}{\sqrt{3}} (\xi_r+\sqrt{2}\chi_r). \label{eq:cust}
\end{align}
It is seen that $H_5^0$ and $\tilde{H}_1$ are CP-even states, whereas $\tilde{H}_3^0$ is a CP-odd state. 
In Eq.~(\ref{eq:cust}), the scalar fields with a tilde are not mass eigenstates in general.  They can in principle
mix with the corresponding scalar fields from the Higgs doublet field. 

The mass of the doubly-charged Higgs boson $H_5^{\pm\pm}$ is
\begin{align}
m_{H_5^{++}}^2=\left(s_H^2\lambda_3 -\frac{3}{2}c_H^2\lambda_5\right)v^2
+c_H^2M_1^2+M_2^2. \label{eq:mchipp}
\end{align}
The mass matrix for the CP-odd Higgs states in the basis of ($\phi_i$, $\tilde{H}_3^0$) 
and that for the singly-charged states in the basis of ($\phi^+$, $\tilde{H}_3^+$, $H_5^+$) are given by
\begin{align}
(M^2)_{\text{CP-odd}}=-\left(\frac{1}{2}\lambda_5 v^2-M_1^2\right)\left(
\begin{array}{cc}
s_H^2& -c_Hs_H\\
-c_Hs_H & c_H^2
\end{array}\right),~
(M^2)_{\pm}
=\left(
\begin{array}{ccc}
\lower3ex\hbox{($M^2)_{\text{CP-odd}}$}&&0\\
&&0\\
0&0&m_{H_5^{++}}^2
\end{array}
\right). \notag
\end{align}
The mass matrix for the CP-even Higgs states in the basis of ($\phi_r$, $\tilde{H}_1^0$, $\tilde{H}_5^0$) is
\begin{align}
(M^2)_{\text{CP-even}}
=
\left(
\begin{array}{ccc}
 (M^2)_{11} & (M^2)_{12} & 0\\
 (M^2)_{12} & (M^2)_{22} & 0\\
 0 & 0 & m_{H_5^{++}}^2 
\end{array}\right), \notag
\end{align}
where the elements of the $2\times 2$ submatrix are
\begin{subequations}
\begin{align}
(M^2)_{11}&=8c_H^2\lambda_1v^2, \\
(M^2)_{22}&=s_H^2(3\lambda_2+\lambda_3)v^2+c_H^2M_1^2
-\frac{1}{2}M_2^2,\\
(M^2)_{12}&=\sqrt{\frac{3}{2}}s_Hc_H[(2\lambda_4+\lambda_5)v^2-M_1^2].
\end{align}
\label{eq:Meven_ele}
\end{subequations}
The mass eigenstates are related to the above-mentioned states via the following unitary transformations
\begin{align}
\left(
\begin{array}{c}
\phi_i\\
\tilde{H}_3^0
\end{array}\right)
=U_{\text{CP-odd}}
\left(
\begin{array}{c}
G^0\\
H_3^0
\end{array}\right),~
\left(
\begin{array}{c}
\phi^\pm\\
\tilde{H}_3^\pm\\
H_5^\pm
\end{array}\right)
=U_\pm
\left(
\begin{array}{c}
G^\pm\\
H_3^\pm\\
H_5^\pm
\end{array}\right),~
\left(
\begin{array}{c}
\phi_r\\
\tilde{H}_1^{0}\\
H_5^0
\end{array}\right)
=U_{\text{CP-even}}
\left(
\begin{array}{c}
h\\
H_1^0\\
H_5^0
\end{array}\right), 
\end{align}
where $G^\pm$ and $G^0$ are the NG bosons for the longitudinal components of the $W^\pm$ and $Z$ bosons. 
The explicit forms of the unitary matrices are
\begin{align}
U_{\text{CP-odd}}=
\left(
\begin{array}{cc}
c_H & -s_H \\
s_H & c_H
\end{array}\right),~ 
U_\pm=\left(
\begin{array}{ccc}
\lower3ex\hbox{$U_{\text{CP-odd}}$}&&0\\
&&0\\
0&0&1
\end{array}
\right)
,~ U_{\text{CP-even}}=\left(
\begin{array}{ccc}
 c_\alpha & -s_\alpha &0\\
 s_\alpha & c_\alpha &0\\
0 & 0 & 1 
\end{array}\right), 
\end{align}
where $c_\alpha=\cos\alpha$, $s_\alpha=\sin\alpha$ and the mixing angle $\alpha$ is defined by
\begin{align}
\tan2\alpha =\frac{2(M^2)_{12}}{(M^2)_{11}-(M^2)_{22}}. 
\end{align}
The masses of the singly-charged Higgs bosons ($H_5^\pm$ and $H_3^\pm$),  
the CP-odd Higgs boson ($H_3^0$) and the CP-even Higgs bosons ($H_5^0$, $H_1^0$ and $h$) are then
\begin{align}
&m_{H_5^+}^2=m_{H_5^0}^2=m_{H_5^{++}}^2,\quad m_{H_3^+}^2=m_{H_3^0}^2=-\frac{1}{2}\lambda_5v^2+M_1^2,\notag\\
&m_{h}^2=(M^2)_{11}c_\alpha^2+(M^2)_{22}s_\alpha^2
+2(M^2)_{12}s_\alpha c_\alpha,\notag\\
&m_{H_1^0}^2=(M^2)_{11}s_\alpha^2+(M^2)_{22}c_\alpha^2
-2(M^2)_{12}s_\alpha c_\alpha.\label{eq:mass}
\end{align}
It is observed that $H_5^{\pm\pm}$, $H_5^\pm$ and $H_5^0$ are degenerate in mass 
and so are $H_3^\pm$ and $H_3^0$ because of the custodial invariance in the Higgs potential. 
Therefore, the Higgs boson masses can be conveniently written as
\begin{align}
m_{H_5}^2\equiv m_{H_5^{++}}^2=m_{H_5^{+}}^2=m_{H_5^{0}}^2,\quad 
m_{H_3}^2\equiv m_{H_3^{+}}^2=m_{H_3^{0}}^2,\quad 
m_{H_1}^2\equiv m_{H_1^{0}}^2. 
\end{align}
The five dimensionless couplings in the potential, $\lambda_1, \dots, \lambda_5$, can be substituted by
the five physical parameters $m_{H_5}$, $m_{H_3}$, $m_{H_1}$, $m_{h}$ and $\alpha$ as follows: 
\begin{align}
\lambda_1 &=\frac{1}{8v^2c_H^2}(m_h^2c_\alpha^2+m_{H_1}^2s_\alpha^2),\notag\\
\lambda_2 &=\frac{1}{6v^2s_H^2}
\left[2m_{H_1}^2c_\alpha^2+2m_h^2s_\alpha^2+3M_2^2-2m_{H_5}^2+6c_H^2(m_{H_3}^2-M_1^2)\right],\notag\\
\lambda_3 &= \frac{1}{v^2s_H^2}\left[c_H^2(2M_1^2-3m_{H_3}^2)+m_{H_5}^2-M_2^2\right],\notag\\
\lambda_4&= \frac{1}{6v^2s_Hc_H}\left[\frac{\sqrt{6}}{2}s_{2\alpha}(m_h^2-m_{H_1}^2)+3s_Hc_H(2m_{H_3}^2-M_1^2)\right],\notag\\
\lambda_5&= \frac{2}{v^2}(M_1^2-m_{H_3}^2). \label{lambdas}
\end{align}

The decoupling limit of this model can be obtained when we take the $v_\Delta\to 0$ limit (or equivalently $s_H\to 0$). 
In this limit, the mass formulae of the Higgs bosons reduce to
\begin{align}
m_{H_5}^2 =  -\frac{3}{2}\lambda_5v^2 + M_1^2 + M_2^2,~
m_{H_3}^2 =  -\frac{1}{2}\lambda_5v^2 + M_1^2 ,~
m_{H_1}^2 =  M_1^2 -\frac{1}{2}M_2^2,~
m_h^2     = 8\lambda_1 v^2. \label{dec}
\end{align}
Notice that $M_2^2$ is proportional to $s_H\mu_2$, and thus it becomes zero in this limit for a fixed value of $\mu_2$. 
If one wants to fix $M_2^2$ at a finite value, $\mu_2$ has to be taken to infinity to compensate $s_H\to 0$ and eventually violates perturbativity in this model.  Therefore, $M_2^2=0$ is the natural choice in this limit. 
On the other hand, $M_1^2$ is proportional to $\mu_1/s_H$. 
Even in the $s_H\to 0$ limit, we can take a finite value for $M_1^2$ as long as $\mu_1 \to 0$ at the same rate as $s_H$.  
Consequently, the triplet-like Higgs bosons decouple when $M_1^2\gg v^2$, and 
only $h$ remains at the electroweak scale and acts like the SM Higgs boson. 
In addition, in the decoupling region $v_\Delta\simeq 0$, we find a simple mass relation for the 
triplet-like Higgs bosons:
\begin{align}
m_{H_1}^2= \frac{3}{2}m_{H_3}^2-\frac{1}{2}m_{H_5}^2.  \label{m1_pred}
\end{align}

For the convenience in discussing interactions between leptons and the Higgs triplet field, 
we reorganize the Higgs fields as follows: 
\begin{align}
\phi=\left(
\begin{array}{c}
\phi^+\\
\phi^0
\end{array}\right),\quad
\chi=\left(
\begin{array}{cc}
\frac{\chi^+}{\sqrt{2}} & -\chi^{++} \\
\chi^0 & -\frac{\chi^+}{\sqrt{2}} 
\end{array}\right),\quad 
\xi=\left(
\begin{array}{cc}
\frac{\xi^0}{\sqrt{2}} & -\xi^+ \\
\xi^- & -\frac{\xi^0}{\sqrt{2}}
\end{array}\right). \label{22mat}
\end{align}
The relationship between the two representations in Eqs.~(\ref{eq:Higgs_matrices}) and (\ref{22mat}) are given in Appendix~A.  With the introduction of the $\chi$ field above, the Yukawa interactions between the lepton doublets and the Higgs triplet are
\begin{align}
\mathcal{L}_{\nu}&=h_{ij}\overline{L_L^{ic}}i\tau_2\chi L_L^j+\text{h.c.} \label{Eq:nuYukawa}
\end{align}
If we assign two units of lepton number to $\chi$, then the $\lambda_5$ and $\mu_1$ terms in the Higgs potential violate the 
lepton number.  If we then take $\lambda_5=\mu_1=0$, $H_3^0$ becomes massless and corresponds to the NG boson for the spontaneous breakdown of the global $U(1)$ lepton number symmetry.  In fact, $H_3^\pm$ are also massless in that case because of the custodial symmetry. 

The Majorana mass of neutrinos is derived as
\begin{align}
(m_{\nu})_{ij}&=h_{ij}v_\Delta=\frac{h_{ij}}{2\sqrt{2}}vs_H. 
\end{align}
This mass matrix can be diagonalized by the Pontecorvo-Maki-Nakagawa-Sakata matrix $V_{\text{PMNS}}$,  
and the Yukawa matrix $h_{ij}$ can be rewritten as
\begin{align}
h_{ij}=2\sqrt{2}\frac{V_{\text{PMNS}}^Tm_\nu^{\text{diag}}V_{\text{PMNS}}}{vs_H}. 
\end{align}
The left-handed neutrino fields are then transformed as
\begin{align}
\nu_L = V_{\text{PMNS}}^\dagger\nu_L'.
\end{align}
For simplicity, we hereafter assume that $V_{\text{PMNS}}$ is the unit matrix and the mass eigenvalues of 
$m_\nu^{\text{diag}}$ are degenerate: $m_\nu^{\text{diag}}=\text{diag}(m_\nu,m_\nu,m_\nu)$. 
In terms of the scalar mass eigenstates, the interaction terms are
\begin{align}
\mathcal{L}_\nu 
&=\frac{2\sqrt{2}m_\nu}{s_Hv}H_5^{++}\overline{e_i^c}P_Le_i
-\frac{2\sqrt{2}m_\nu}{s_Hv}\left(H_5^++c_HH_3^++s_HG^+\right)\overline{\nu_i^c}P_Le_i\Big]\notag\\
&+\frac{2m_\nu}{s_Hv}\left[\frac{1}{\sqrt{3}}(H_5^0+\sqrt{2}s_\alpha h+c_\alpha H_1^0)+i(G^0 s_H +H_3^0c_H)\right]\overline{\nu_i^c}P_L\nu_i
+\text{h.c.}\label{nY}
\end{align}

The Yukawa interaction between the fermions of one generation and the Higgs doublet $\phi$ is given by
\begin{align}
\mathcal{L}_{Y}&=-Y_u\overline{Q_L}\tilde{\phi}u_R-Y_d\overline{Q_L}\phi d_R
-Y_e\overline{L_L}\phi e_R+\text{h.c.}, \label{Yukawa}
\end{align}
with $\tilde{\phi}=i\tau_2 \phi^*$. 
In terms of the fermion masses $m_{f}=\frac{vc_H}{\sqrt{2}}Y_f$ and the physical Higgs states,
the interaction terms are expressed as 
\begin{align}
\mathcal{L}_{Y}&=
-\sum_{f=u,d,e}\frac{m_f}{v}
\left[\frac{c_\alpha}{c_H}\bar{f}fh-\frac{s_\alpha}{c_H}\bar{f}fH_1^0+i\text{Sign}(f) \tan\theta_H\bar{f}\gamma_5fH_3^0\right]
\notag\\
&-\frac{\sqrt{2}V_{ud}}{v}\left[\tan\theta_H\bar{u}(m_u P_L-m_d P_R)dH_3^+\right]+\frac{\sqrt{2}m_e}{v}\tan\theta_H\bar{\nu}P_ReH_3^++\text{h.c.},  \label{Yukawa1}
\end{align}
where $V_{ud}$ is one element of the Cabibbo-Kobayashi-Maskawa (CKM) matrix, Sign$(f=u)=+1$ and Sign$(f=d,e)=-1$.

Finally, we discuss the kinetic terms for the Higgs fields
\begin{align}
\mathcal{L}_{\text{kin}}&=\frac{1}{2}\text{tr}(D_\mu \Phi)^\dagger (D^\mu \Phi)
+\frac{1}{2}\text{tr}(D_\mu \Delta)^\dagger (D^\mu \Delta), 
\end{align}
where the covariant derivatives are
\begin{align}
D_\mu \Phi =\partial_\mu\Phi+ig\frac{\tau^a}{2}W_\mu^a\Phi-ig'B_\mu \Phi\frac{\tau^3}{2},\\
D_\mu \Delta =\partial_\mu\Delta+igt^aW_\mu^a\Delta-ig'B_\mu \Delta t^3. 
\end{align}
The masses of the gauge bosons are obtained under the condition of $v_\chi=v_\xi \equiv v_\Delta$ as 
\begin{align}
m_W^2 = \frac{g^2}{4}v^2,\quad m_Z^2=\frac{g^2}{4\cos^2\theta_W}v^2.
\end{align}
Thus, the electroweak rho parameter $\rho = m_W^2/(m_Z^2\cos^2\theta_W)$ is unity at the tree level. 
One-loop corrections to $\rho$ have been calculated in Ref.~\cite{GVW2} for the GM model. 
The deviation of $\rho$ from unity depends on the logarithm of the triplet-like Higgs boson masses and, 
therefore, the one-loop effect is not important in this model.

The Gauge-Gauge-Scalar (Gauge-Scalar-Scalar) vertices
are listed in Table~\ref{GGS} (Table~\ref{GSS}) in Appendix~B. 
We note that there is the $H_5^\pm W^\mp Z$ vertex at the tree level in the GM model (see Table~\ref{GGS}).  In the Higgs-extended models with $\rho=1$ at the tree level and having singly-charged Higgs bosons ({\it e.g.}, the 2HDM), the $H^\pm W^\mp Z$ vertex is absent at the tree level~\cite{Grifols} and can only be induced at loop levels.  Therefore, the magnitude of this vertex in such models is much smaller than that in the GM model.  Thus, this vertex can be used to discriminate models with singly-charged Higgs bosons.  The possibility of measuring the $H^\pm W^\mp Z$ vertex has been discussed in Refs.~\cite{HWZ-LHC} for the LHC
and in Ref.~\cite{HWZ-ILC} for future linear colliders.

\section{Constraints \label{sec:constraints}}
In this section, we discuss constraints on the parameter space of the GM model. 
First, we consider the theoretical constraints from perturbative unitarity and vacuum stability. 
Secondly, as experimental constraints, we consider the $Zb\bar{b}$ data and other $B$ physics data. 

\subsection{Perturbative unitarity and vacuum stability bounds}

\begin{figure}[t]
\begin{center}
\includegraphics[width=50mm]{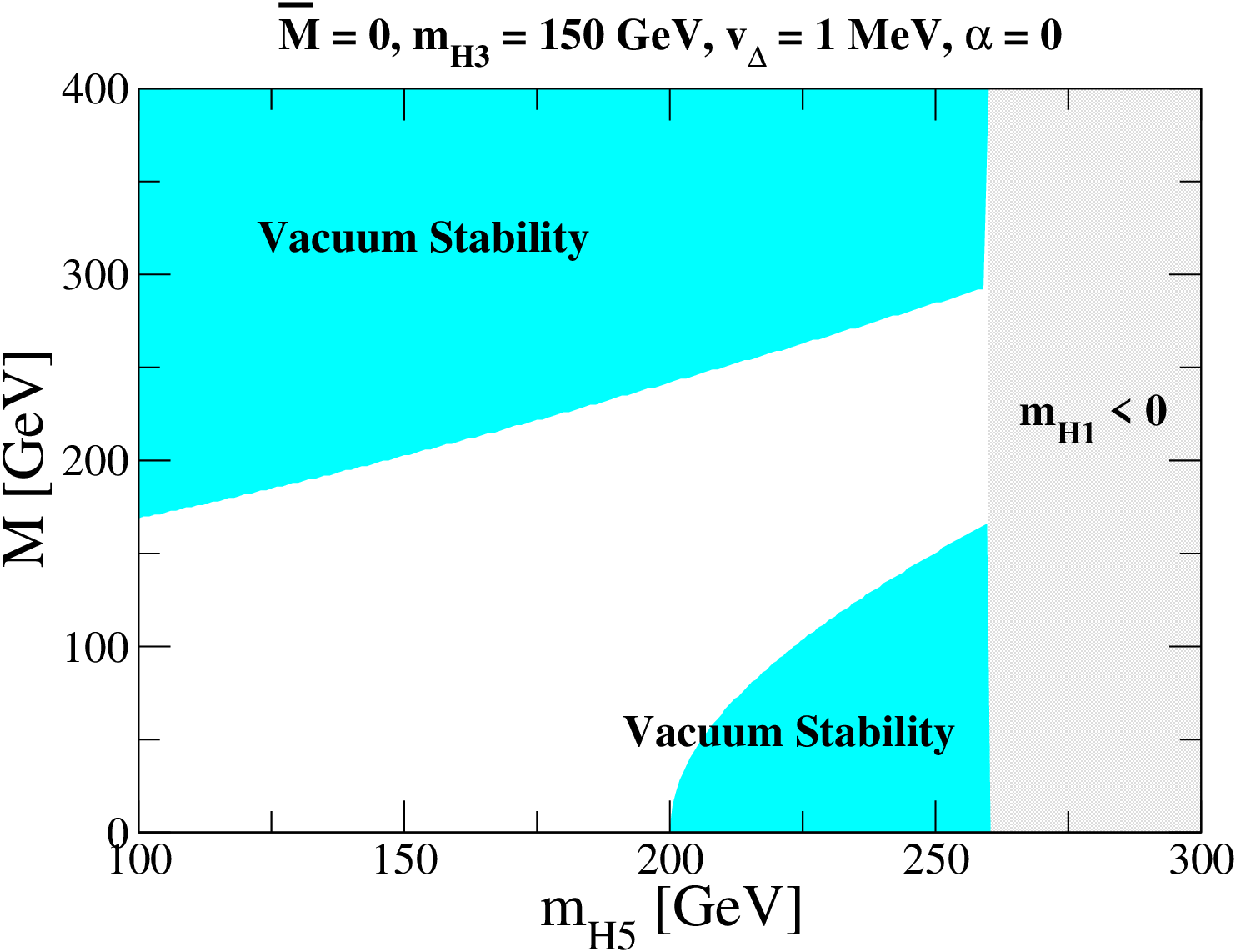}\hspace{3mm}
\includegraphics[width=50mm]{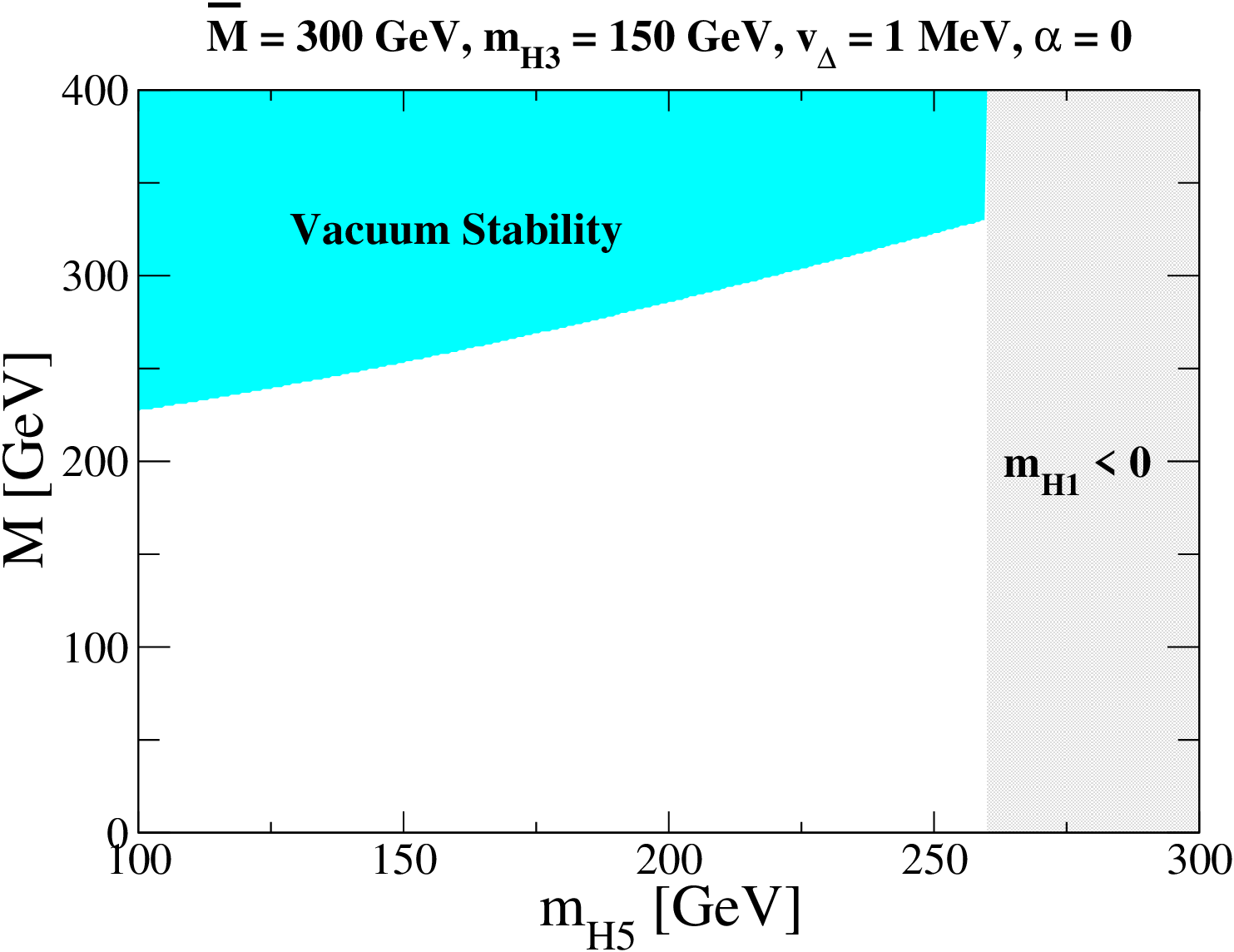}\hspace{3mm}
\includegraphics[width=50mm]{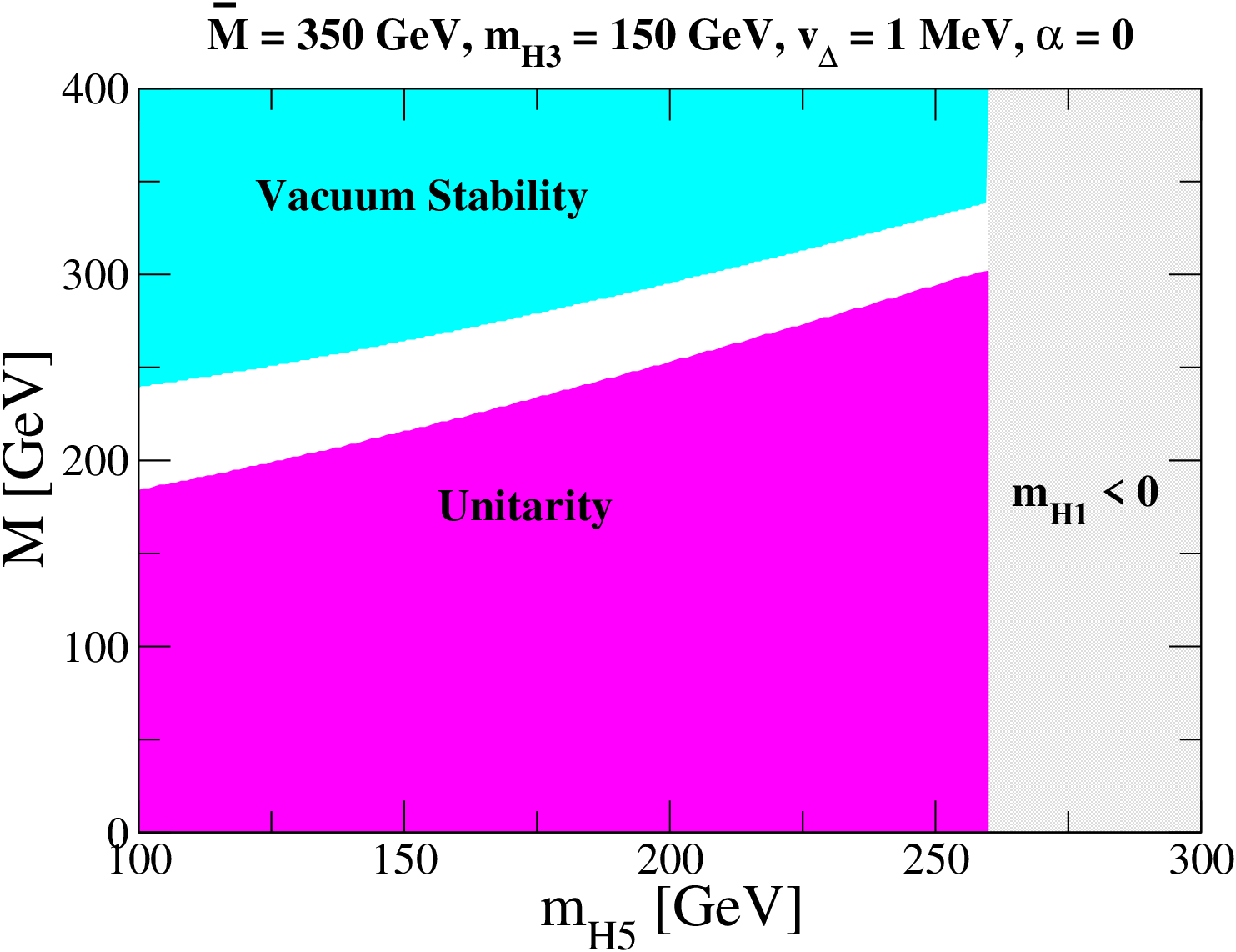}
\end{center}
\caption{Constraints from the unitarity and vacuum stability in the $M$-$m_{H_5}$ plane. 
In all the plot, the uncolored regions are allowed, and the 3-plet Higgs mass is taken to be 150 GeV, $v_\Delta =1$ MeV and 
$\alpha=0$. 
Blue, gray and pink shaded regions are respectively excluded by the vacuum stability bound, unitarity bound and a
negative singlet Higgs mass ($m_{H_1}<0$). 
The left, center and right plot show the case of $\bar{M}=0$, 300 GeV and 350 GeV, respectively. 
}
\label{theory_const1}
\begin{center}
\includegraphics[width=50mm]{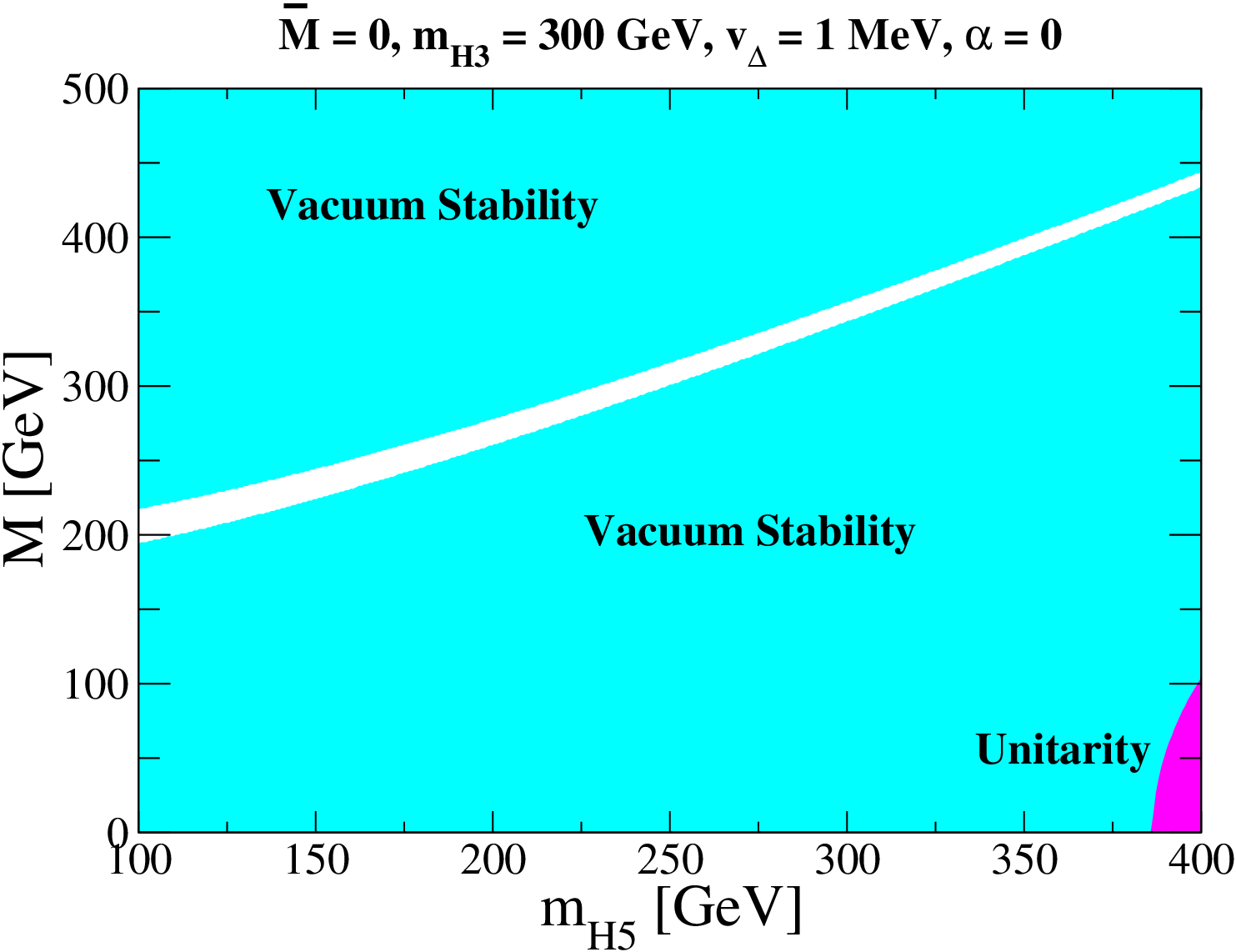}\hspace{3mm}
\includegraphics[width=50mm]{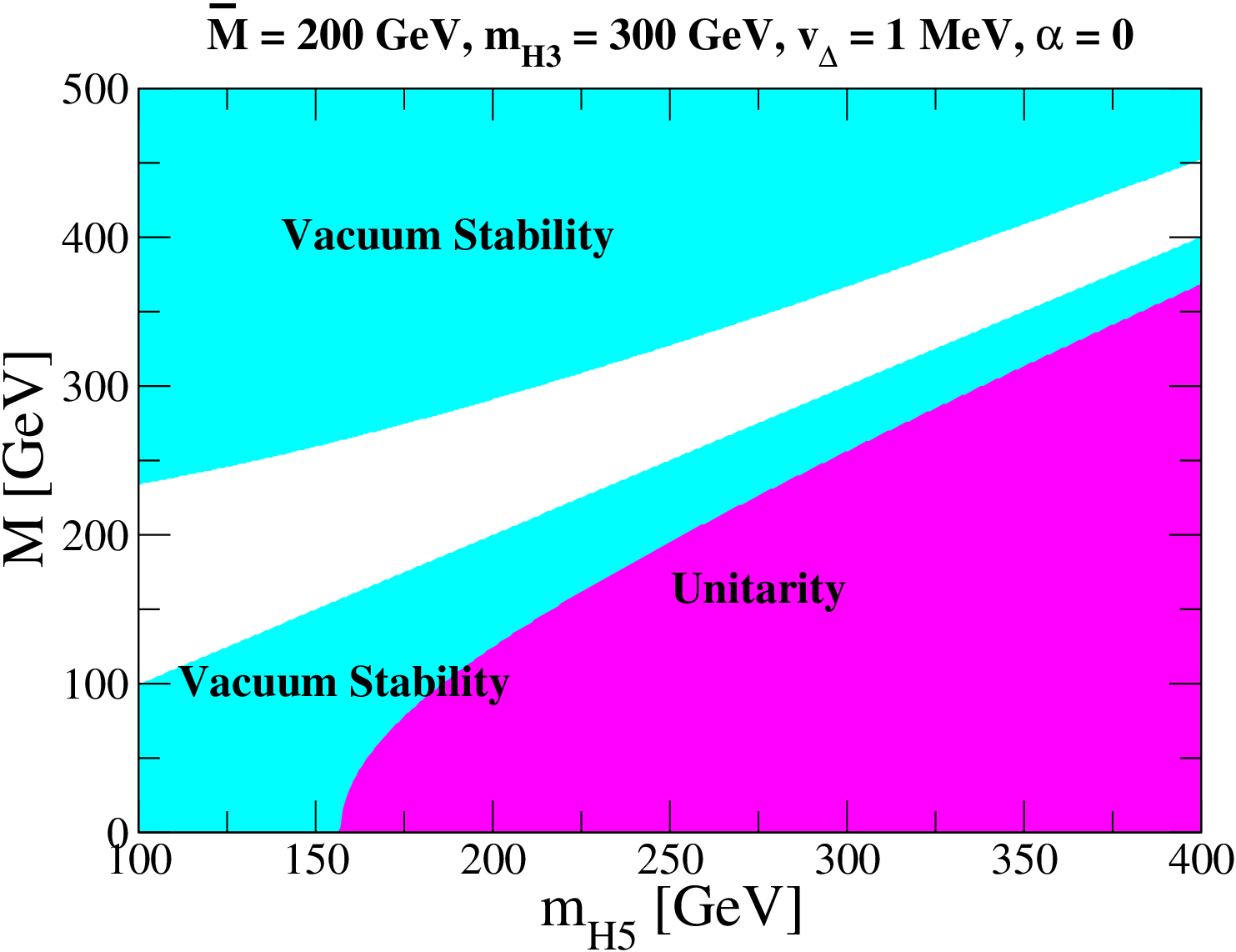}\hspace{3mm}
\includegraphics[width=50mm]{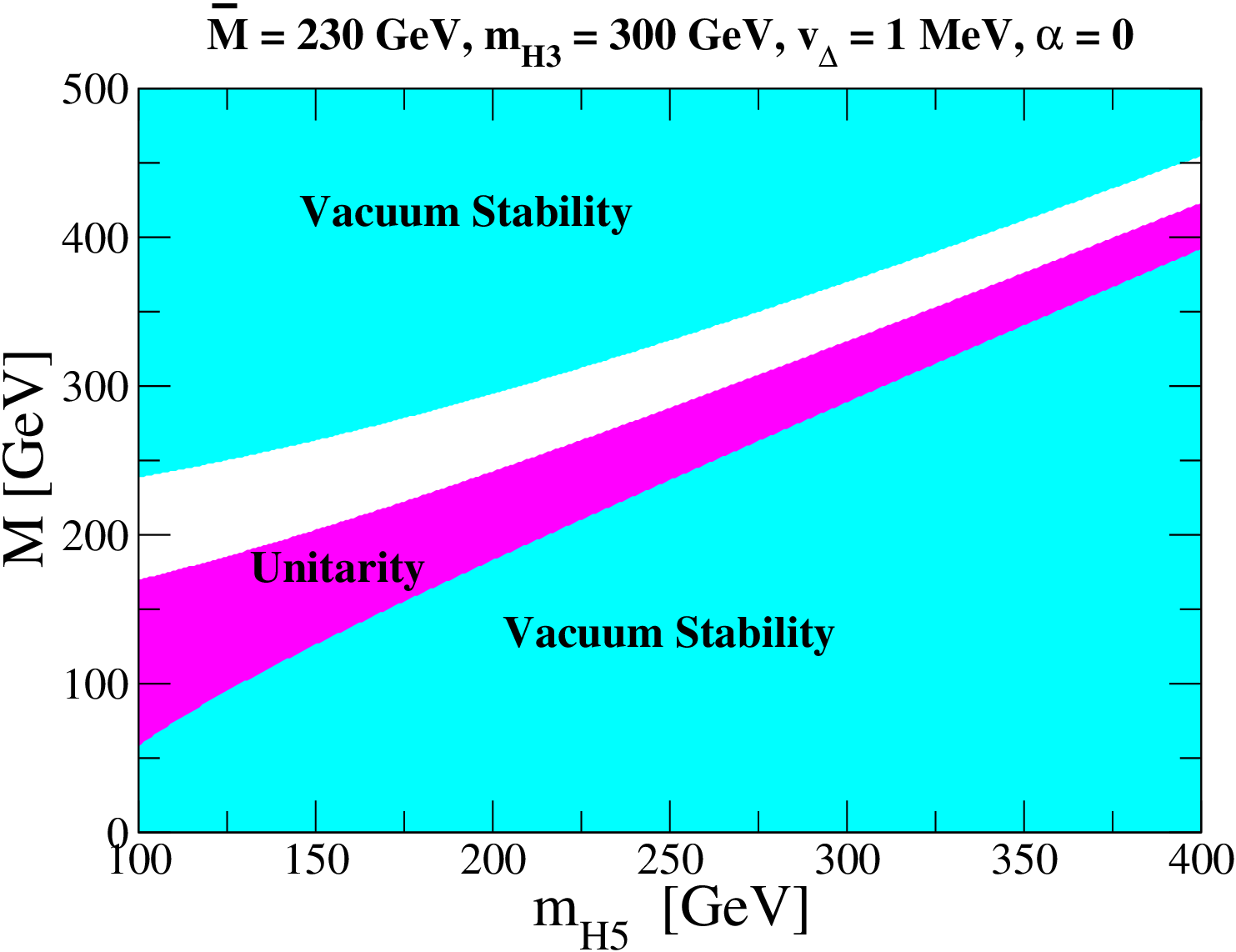}
\end{center}
\caption{Constraints from the unitarity and vacuum stability in the $M$-$m_{H_5}$ plane. 
In all the plot, the uncolored regions are allowed, and the 3-plet Higgs mass is taken to be 300 GeV, $v_\Delta =1$ MeV and 
$\alpha=0$. 
Blue and pink shaded regions are respectively excluded by the vacuum stability bound and the unitarity bound. 
The left, center and right plot show the case of $\bar{M}=0$, 200 GeV and 230 GeV, respectively.  }
\label{theory_const2}
\end{figure}

The perturbative unitarity bound for the GM model has been studied in Ref.~\cite{Aoki_Kanemura} and can be directly applied to our analysis.  Before doing so, we will make a change in the parameterization.  This is because Eqs.~(\ref{lambdas}) suggest apparent divergences in $\lambda_{2,3,4}$ in the limit $v_\Delta \ll v$.  However, this is only an artefact that can be avoided by reparameterization.

We therefore select the following parameterization
\begin{align}
m_{H_1}^2 = \frac{1}{2}\left(3m_{H_3}^2-m_{H_5}^2+3s_H^2 \bar{M}^2\right)
,\quad M_1^2=\frac{1}{2}\left(3m_{H_3}^2-m_{H_5}^2+M^2\right),\quad M_2^2=M^2,
\label{mass_set}
\end{align}
in terms of which all the dimensionless couplings can be rewritten for $\sin\alpha =0$ as
\begin{align}
\lambda_1 &=\frac{m_h^2}{8v^2c_H^2},\quad \lambda_2 = \frac{m_{H_3}^2-m_{H_5}^2+M^2+\bar{M}^2}{2v^2},\quad
\lambda_3 = \frac{m_{H_5}^2-M^2}{v^2},\notag\\
\lambda_4 &= \frac{m_{H_3}^2+m_{H_5}^2-M^2}{4v^2},\quad \lambda_5=\frac{m_{H_3}^2-m_{H_5}^2+M^2}{v^2}.
\end{align}
It is seen that the $v_\Delta$ dependence drops out in $\lambda_{2,3,4}$ and no divergent $\lambda$'s appear even when $v_\Delta\ll v$.  

For the vacuum stability condition, we require that the potential is bounded from below in any direction with large scalar fields. 
This condition imposes constraints on the dimensionless coupling constants $\lambda_1, \dots, \lambda_5$. 
In the GM model, we then derive the following inequalities 
\begin{align}
&\lambda_1>0,\quad \lambda_2+\lambda_3>0,\quad \lambda_2+\frac{1}{2}\lambda_3>0,\quad -|\lambda_4|+2\sqrt{\lambda_1(\lambda_2+\lambda_3)}>0,\notag\\
&\lambda_4-\frac{1}{4}|\lambda_5|+\sqrt{2\lambda_1(2\lambda_2+\lambda_3)}>0.\label{vs}
\end{align}
They have taken into account the positivity of all combinations of two non-zero scalar fields, as have been discussed in Ref.~\cite{Arhrib_VS} for the HTM.  

Fig.~\ref{theory_const1} shows the regions excluded by the unitarity and the vacuum stability constraints for the case of $m_{H_3}=150$ GeV and $v_\Delta=1$ MeV.  
The left, center and right plots show the cases for $\bar{M}$=0, 300, 350 GeV, respectively. 
For the unitarity bound, we consider the $S$-wave amplitudes for elastic scatterings of two scalar boson states and require their absolute values of the eigenvalues to be less than 1.
It is observed that the allowed regions by the unitarity bound for larger $\bar{M}$
is smaller than those for smaller $\bar{M}$.  This is because the $\lambda_2$ coupling increases as $\bar{M}$ becomes larger. 
In fact, the excluded regions are determined by the following unitarity condition~\cite{Aoki_Kanemura};
\begin{align}
\left|12\lambda_1+22\lambda_2+14\lambda_3\pm\sqrt{(12\lambda_1-22\lambda_2-14\lambda_3)^2+144\lambda_4^2}\right|<16\pi. 
\end{align}
On the other hand, the vacuum stability bound becomes milder as $\bar{M}$ is taken to be a larger value because of the increasing $\lambda_2$ coupling. 
For a fixed value of $m_{H_5}$ and $\bar{M}$, a larger $M$ value is allowed (excluded) by the unitarity (vacuum stability) bound. 

Fig.~\ref{theory_const2} also shows the regions excluded by the unitarity and the vacuum stability conditions for the case of $m_{H_3}=300$ GeV and $v_\Delta=1$ MeV. 
The allowed regions are much smaller than those in the case of $m_{H_3}=150$ GeV. 
The excluded regions from the vacuum stability for smaller (larger) values of $M$ are determined by the third (fourth) inequality in Eq.~(\ref{vs}).

In the case of larger $v_\Delta$ values ({\it e.g.}, $v_\Delta\gtrsim 10$ GeV), the regions excluded by the unitarity (vacuum stability) condition are larger (smaller) compared to the small $v_\Delta$ case.  This is because the $\lambda_1$ coupling becomes larger.  In addition, the singlet Higgs boson mass gets a larger value, so that the regions excluded due to $m_{H_1}<0$ are smaller in the larger $v_\Delta$ case.

\subsection{$Zb\bar{b}$ data}

The renormalized $Zb\bar{b}$ vertex is defined by~\cite{Haber_Logan}
\begin{align}
\mathcal{L}_{Zb\bar{b}}&=-\frac{e}{s_Wc_W}Z_\mu \bar{b}\gamma^\mu 
(\bar{g}_b^LP_L+\bar{g}_b^RP_R)b,\notag
\end{align}
where the renormalized coupling $\bar{g}_b^{L,R}$ can be expressed as
\begin{align}
\bar{g}_b^{L,R}&=g_b^{L,R}+\delta g_b^{L,R\text{ (SM)}}+\delta g_b^{L,R\text{ (GM)}} \text{ with}\notag\\
g_b^L&=I_b-s_W^2Q_b,\quad g_b^R=-s_W^2Q_b, \label{ren_zbb}
\end{align}
where $\delta g_b^{L,R\text{ (SM)}}$ 
($\delta g_b^{L,R\text{ (GM)}}$) denote the one-loop corrections to the $Zb\bar{b}$ vertices from the SM (GM) contributions,  
where the $W$ boson and the NG boson ($H_3^\pm$) are running in the loop, 
$I_f$ ($Q_f$) is the third component of the isospin (the electric charge) for the field $f$,
and $s_W=\sin\theta_W$ and $c_W^2 = 1-s_W^2$. 
The analytic formulas for $\delta g_b^{L,R\text{ (SM)}}$ is given in Ref.~\cite{Lynn}, 
and their numerical values are calculated as~\cite{Field}
\begin{align}
\delta g_b^{L\text{ (SM)}}=-0.4208,\quad \delta g_b^{R\text{ (SM)}}=0.0774.
\end{align}
The one-loop correction $\delta g_b^{L\text{(GM)}}$ is given in terms of the 
Passarino-Veltman function~\cite{PV} by
\begin{align}
&\delta g_b^{L\text{(GM)}}=
-\frac{e}{s_Wc_W}\frac{2\tan^2\theta_Hm_t^2}{v^2}\frac{1}{16\pi^2}\notag\\
&\times\Big[c_{2W}C_{24}(m_b^2,m_Z^2,m_b^2,m_t,m_{H_3},m_{H_3})
+2s_W^2Q_tC_{24}(m_b^2,m_Z^2,m_b^2,m_{H_3},m_t,m_t)-\frac{1}{2}s_W^2Q_t\notag\\
&+m_t^2(I_t-s_W^2Q_t)C_0(m_b^2,m_Z^2,m_b^2,m_{H_3},m_t,m_t)
-(I_b-s_W^2Q_b)B_1(m_b^2,m_t,m_{H_3})\Big]. 
\end{align}
On the other hand, $\delta g_b^{R\text{(GM)}}$ can be neglected because the corrections are proportional 
to the bottom quark mass [see Eq.~(\ref{Yukawa1})]. 
We can also neglect the contributions from $H_3^0$ loop diagrams for the same reason. 
The renormalized couplings $\bar{g}_b^{L,R}$ can be compared to the experimental value of $R_b^{\text{exp}}$~\cite{pdg}
\begin{align}
R_b^{\text{exp}}=0.21629\pm 0.00066. \label{Rb_exp}
\end{align}

\begin{figure}[t]
\begin{center}
\includegraphics[width=80mm]{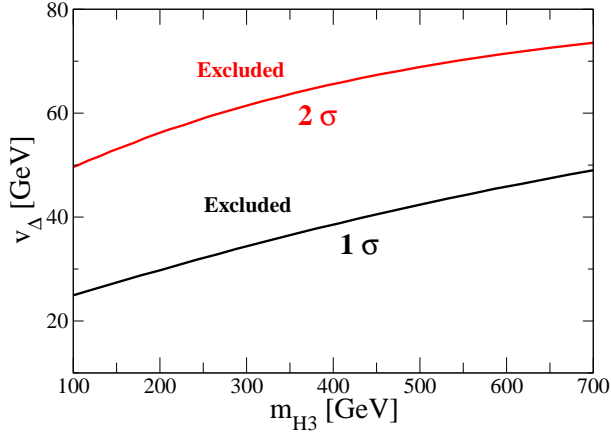}
\end{center}
\caption{Constraint from the $R_b$ data given in Eq.~(\ref{Rb_exp}) on $v_\Delta$ as a function of $m_{H_3}$.  
The region above the black (red) line is excluded at 1$\sigma$ (2$\sigma$) level. }
\label{bound_Zbb}
\end{figure}

In Fig.~\ref{bound_Zbb}, we show the excluded parameter space in the $m_{H_3}$-$v_\Delta$ plane using the $R_b$ data in Eq.~(\ref{Rb_exp}).  Basically, the upper bound on $v_\Delta$ increases monotonically with $m_{H_3}$.  The 2$\sigma$ bound is about 25 GeV more relaxed than the 1$\sigma$ bound over the considered range.
We note in passing that the constraint of the $b\to s\gamma$ data for the GM model 
is similar to that in the Type-I 2HDM~\cite{barger, typeX} and is milder than the $R_b$ constraint.

\section{Higgs decays \label{sec:decays}}

\begin{figure}[!t]
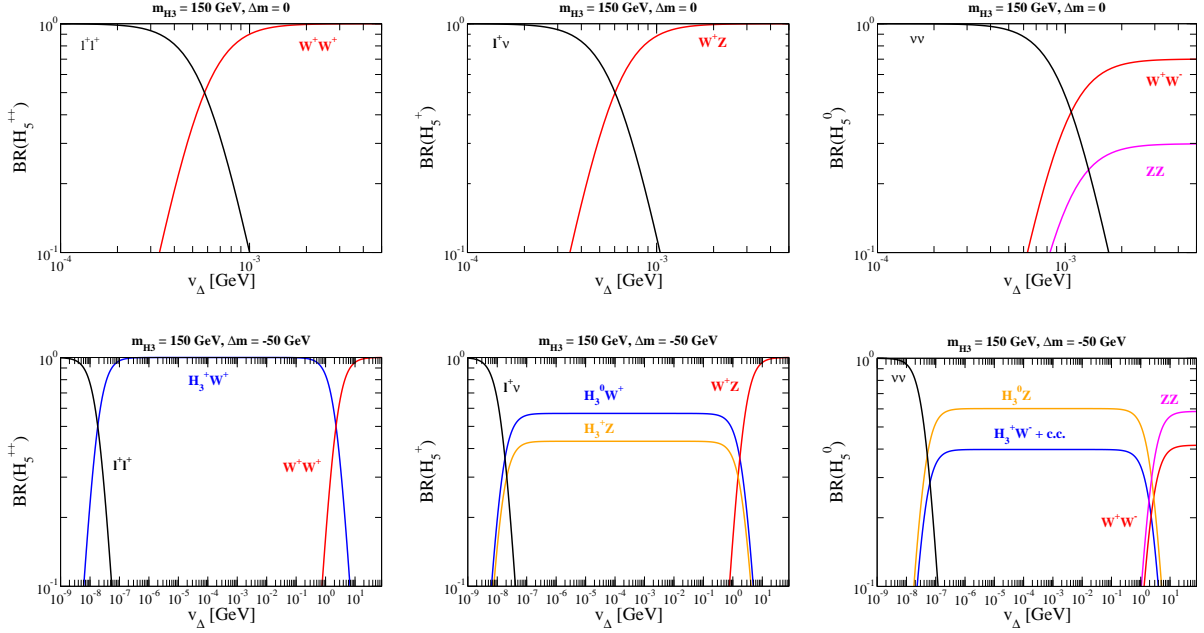

\begin{center}
\includegraphics[width=50mm]{BR_H5pp_dm0_vd_150.eps}\hspace{3mm}
\includegraphics[width=50mm]{BR_H5p_dm0_vd_150.eps}\hspace{3mm}
\includegraphics[width=50mm]{BR_H50_dm0_vd_150.eps}\\\vspace{5mm}
\includegraphics[width=50mm]{BR_H5pp_dmm50_vd_150.eps}\hspace{3mm}
\includegraphics[width=50mm]{BR_H5p_dmm50_vd_150.eps}\hspace{3mm}
\includegraphics[width=50mm]{BR_H50_dmm50_vd_150.eps}
\end{center}
\caption{The decay branching ratios of $H_5^{++}$ (left),  $H_5^+$ (center) and $H_5^0$ (right) as a function of $v_\Delta$.  
We take $m_{H_3}=150$ GeV, $m_h$=125 GeV and $\alpha =0$ in all the plots.  
The mass difference $\Delta m$ is taken to be 0 for the upper three plots, 
and to be $-50$ GeV for the lower three plots. }
\label{decay_150_vd_2}
\end{figure}

In this section, we discuss the decay of the triplet-like Higgs bosons, namely 
the 5-plet Higgs bosons $H_5$ ($=H_5^{\pm\pm},H_5^\pm$ or $H_5^0$), 3-plet Higgs bosons 
$H_3$ ($=H_3^\pm$ or $H_3^0$) and the singlet Higgs boson $H_1^0$. 
Decay branching ratios of the Higgs bosons depend on the mass parameters $m_{H_5}$, $m_{H_3}$ and $m_{H_1}$, 
the VEV of the triplet field $v_\Delta$, and the mixing angle $\alpha$. 
The mass of the SM-like Higgs boson $h$ is fixed at 125 GeV. 
Using the mass relation given in Eq.~(\ref{m1_pred}), we can treat
$m_{H_1}$ as an dependent parameter determined by $m_{H_3}$ and $m_{H_5}$. 
Hereafter, we take $\Delta m \equiv m_{H_3}-m_{H_5}$, $m_{H_3}$ and $v_\Delta$ as the input parameters, and assume $\alpha=0$ for simplicity.  
Once we apply the mass relation, there are three different patterns of masses for 
the triplet-like Higgs bosons. 
In the case of $\Delta m=0$, all the masses of the triplet-like Higgs bosons are degenerate: $m_{H_5}=m_{H_3}=m_{H_1}$, whereas in the case of $\Delta m>0$ ($\Delta m<0$), the mass spectrum is then $m_{H_1}>m_{H_3}>m_{H_5}$ ($m_{H_5}>m_{H_3}>m_{H_1}$).

First, we consider the decays of the 5-plet Higgs bosons. 
In the case of $\Delta m \geq 0$, the 5-plet Higgs bosons can decay into weak gauge boson pairs or lepton pairs depending on the magnitude of $v_\Delta$. 
When $\Delta m <0$, the 5-plet Higgs bosons can decay into a 3-plet Higgs boson and a gauge boson, such as $H_5^{++}\to W^+H_3^+$ and $H_5^{+}\to W^+H_3^0$, in addition to the two decay modes allowed in the case of $\Delta m \geq 0$. 

In Fig.~\ref{decay_150_vd_2}, the decay branching ratios of $H_5^{++}$, $H_5^+$ and $H_5^0$ are shown as a function of $v_\Delta$ in the case of $m_{H_3}=150$ GeV. 
When $\Delta m =0$ (upper row), the main decay modes of $H_5^{++}$, $H_5^+$ and $H_5^0$ 
change from $\ell^+\ell^+$, $\ell^+\nu$, and $\nu\nu$ 
to $W^+W^+$, $W^+Z$, and $W^+W^-$ or $ZZ$ at around $v_\Delta=10^{-3}$ GeV, respectively.  Here $H_5^0$ decays more dominantly into $W^+W^-$ than $ZZ$ because of the mass threshold effect.
When $\Delta m =-50$ GeV (lower row) and for the wide range of 
$10^{-8}\lesssim v_\Delta \lesssim 1$ GeV, the main decay mode of $H_5^{++}$ is
$H_3^+W^+$, those of $H_5^+$ are $H_3^+Z$ and $H_3^0W^+$, and those of $H_5^0$ are $H_3^\pm W^\mp$ and $H_3^0Z$.

\begin{figure}[t]
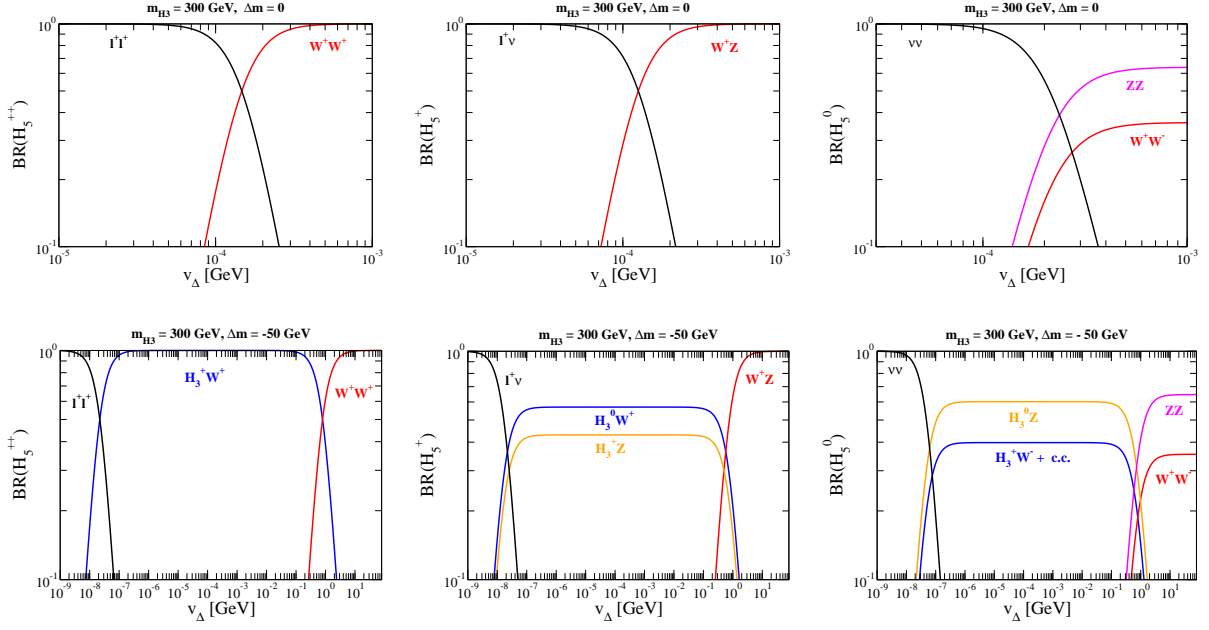

\begin{center}
\includegraphics[width=50mm]{BR_H5pp_dm0_vd_300.eps}\hspace{3mm}
\includegraphics[width=50mm]{BR_H5p_dm0_vd_300.eps}\hspace{3mm}
\includegraphics[width=50mm]{BR_H50_dm0_vd_300.eps}\\\vspace{5mm}
\includegraphics[width=50mm]{BR_H5pp_dmm50_vd_300.eps}\hspace{3mm}
\includegraphics[width=50mm]{BR_H5p_dmm50_vd_300.eps}\hspace{3mm}
\includegraphics[width=50mm]{BR_H50_dmm50_vd_300.eps}
\end{center}
\caption{Same as Fig.~\ref{decay_150_vd_2}, but for $m_{H_3}=300$ GeV.}
\label{decay_300_vd_2}
\end{figure}

Fig.~\ref{decay_300_vd_2} shows the decay branching ratios of the 5-plet Higgs bosons for $m_{H_3}=300$ GeV.
The general behavior here is roughly the same as the case with $m_{H_3}=150$ GeV. 
The crossing point for the main decay modes in each of the upper plots ($\Delta m=0$) slightly shifts to a smaller $v_\Delta$ ($\simeq 10^{-4}$ GeV). 

\begin{figure}[t]
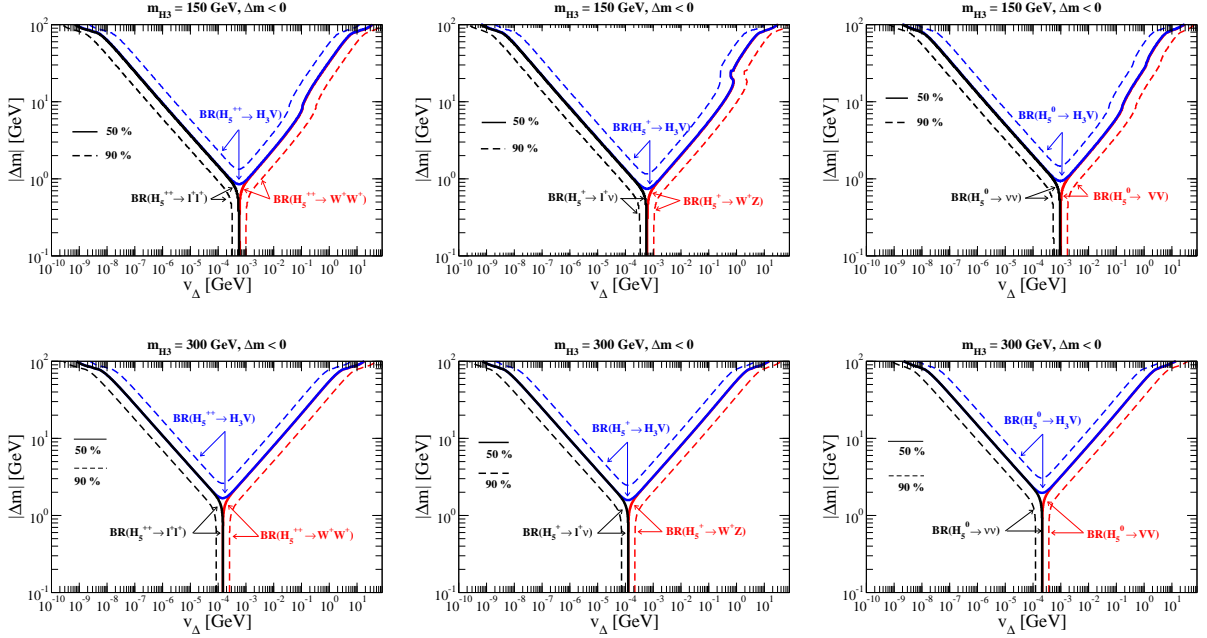

\begin{center}
\includegraphics[width=50mm]{contour_H5pp_150.eps}\hspace{3mm}
\includegraphics[width=50mm]{contour_H5p_150.eps}\hspace{3mm}
\includegraphics[width=50mm]{contour_H50_150.eps}\\\vspace{5mm}
\includegraphics[width=50mm]{contour_H5pp_300.eps}\hspace{3mm}
\includegraphics[width=50mm]{contour_H5p_300.eps}\hspace{3mm}
\includegraphics[width=50mm]{contour_H50_300.eps}
\caption{Contour plots of the decay branching ratios of $H_5^{++}$ (left column),  $H_5^+$ (center column) and $H_5^0$ (right column) on the $v_\Delta$-$|\Delta m|$ plane (with $\Delta m <0$).  We take $m_h$=125 GeV and $\alpha=0$ in all the plots. 
The upper (lower) three plots show the case for $m_{H_3}=150$ GeV (300 GeV).
Each solid (dashed) curve represents the branching ratio of 50\% (90\%) for the corresponding decay mode indicated by the arrow. }
\label{contour3}
\end{center}
\end{figure}

Fig.~\ref{contour3} shows the contour plots of the decay branching ratios of $H_5^{++}$, $H_5^+$ and $H_5^0$ on the $v_\Delta$-$|\Delta m|$ plane (with $\Delta m <0$) for the cases with $m_{H_3}=150$ GeV (upper plots) and $m_{H_3}=300$ GeV (lower plots). 
There are always three distinct regions in this plane. 
In the region of small $|\Delta m|$ and small (large) $v_\Delta$, the main decay modes of the 5-plet Higgs bosons are the a pair of leptons (weak bosons).  In the region of large $|\Delta m|$, they are a 3-plet Higgs boson and a gauge boson, 
denoted by $H_3V$ in the plots ($V=W^\pm$ or $Z$), 
where it is understood that all the possible channels of $H_3V$ should be summed over. 

Secondly, we consider the decays of the 3-plet Higgs bosons. 
The 3-plet Higgs bosons can decay into a pair of fermions through the Yukawa interactions given in Eq.~(\ref{Yukawa1}) and 
a pair of leptons through the neutrino Yukawa interaction given in Eq.~(\ref{nY}), depending on the value of $v_\Delta$ in the case of $\Delta m \simeq 0$.
In the region dominated by fermionic decays, the main decay mode strongly depends on $m_{H_3}$.  When $m_{H_3}$ is smaller than the top quark mass, $H_3^+$ ($H_3^0$) mainly decays into $\tau^+\nu$ or $c\bar{s}$ ($b\bar{b}$), whereas in the case of $m_t<m_{H_3}<2m_t$, 
$H_3^+$ ($H_3^0$) decays into $t\bar{b}$ ($b\bar{b}$). 
Furthermore, when $m_{H_3}$ is larger than $2m_t$, $H_3^0$ decays dominantly into $t\bar{t}$, and $H_3^+$ still mainly into $t\bar{b}$. 
In addition, the 3-plet Higgs bosons can decay into the SM-like Higgs boson $h$ and a gauge boson, {\it e.g.}, $H_3^+ \to hW^+$ and $H_3^0 \to hZ$ if $m_{H_3}$ is larger than $m_h$.  When $\Delta m>0$ ($\Delta m<0$), the 3-plet Higgs bosons can decay into a gauge boson and a 5-plet (singlet) Higgs boson.  

\begin{figure}[t]
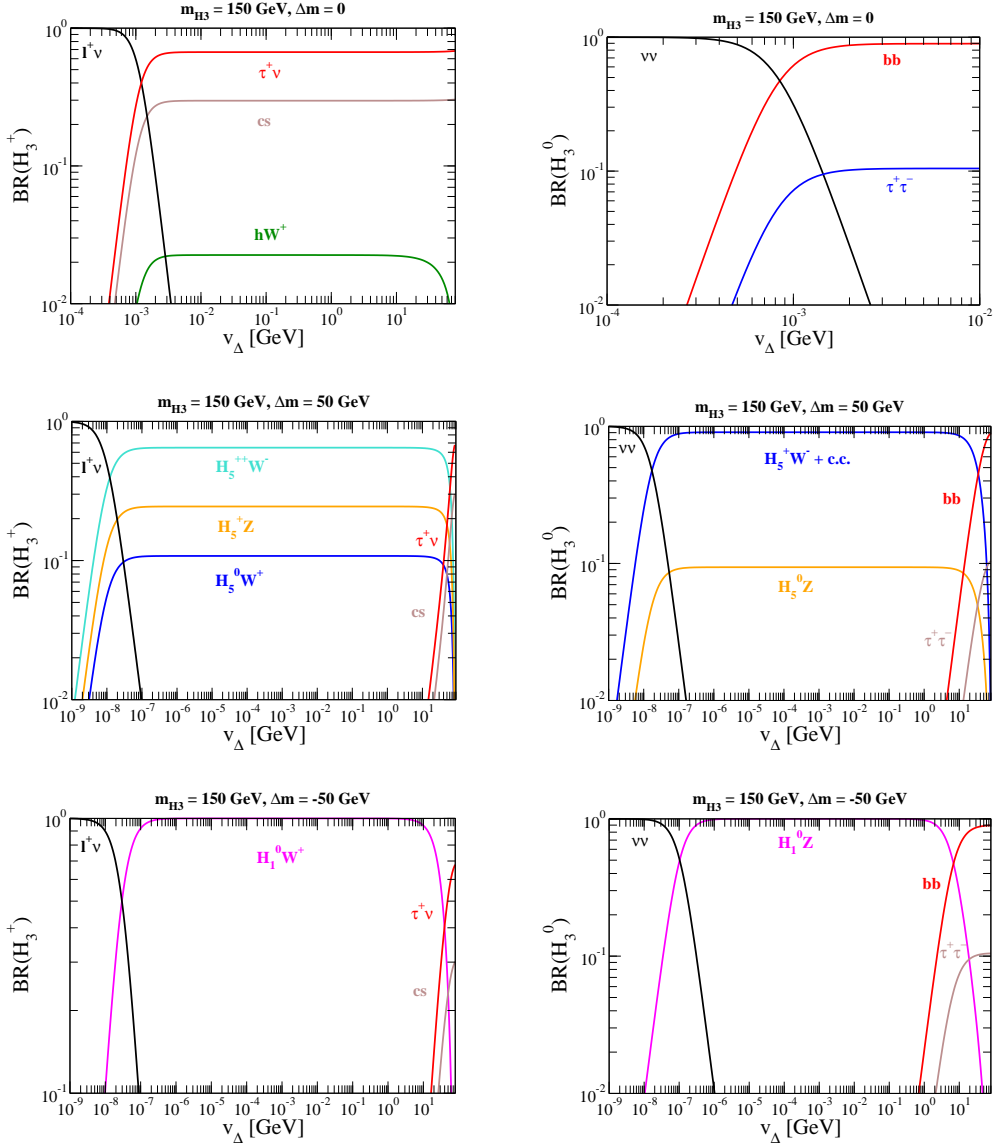

\begin{center}
\includegraphics[width=60mm]{BR_H3p_dm0_vd_150.eps}\hspace{10mm}
\includegraphics[width=60mm]{BR_H30_dm0_vd_150.eps}\\\vspace{5mm}
\includegraphics[width=60mm]{BR_H3p_dm50_vd_150_v3.eps}\hspace{10mm}
\includegraphics[width=60mm]{BR_H30_dm50_vd_150_v2.eps}\\\vspace{5mm}
\includegraphics[width=60mm]{BR_H3p_dmm50_vd_150.eps}\hspace{10mm}
\includegraphics[width=60mm]{BR_H30_dmm50_vd_150.eps}
\end{center}
\caption{Decay branching ratios of $H_3^+$ (left column) and $H_3^0$ (right column) as a function of $v_\Delta$.  We take $m_{H_3}=150$ GeV, $m_h=125$ GeV and $\alpha=0$ in all the plots.  The mass difference $\Delta m$ is fixed to 0 (top plots), 50 GeV (middle plots) 
and $-50$ GeV (bottom plots), respectively.  }
\label{decay_150_vd}
\end{figure}

\begin{figure}[!t]
\begin{center}
\includegraphics[width=60mm]{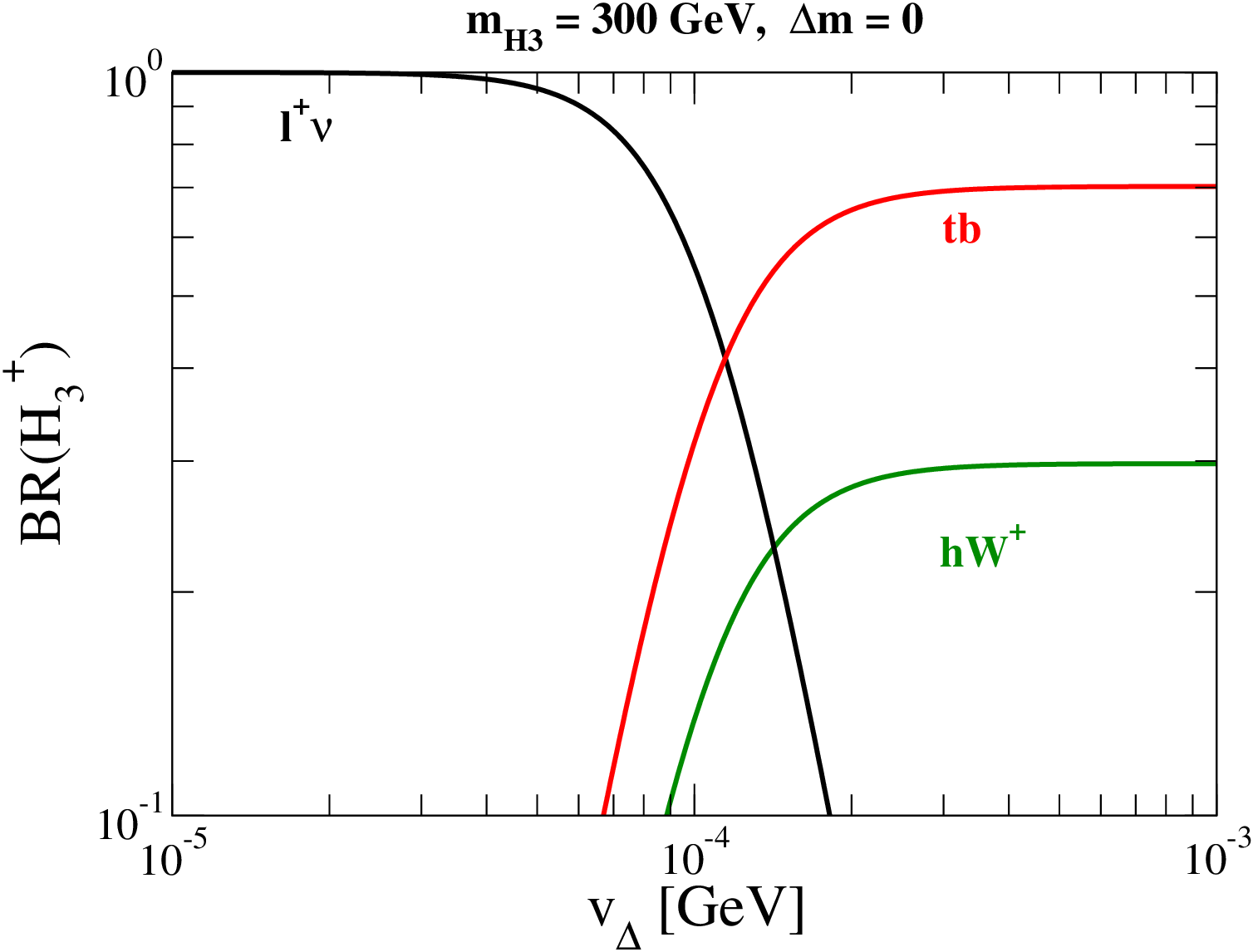}\hspace{10mm}
\includegraphics[width=60mm]{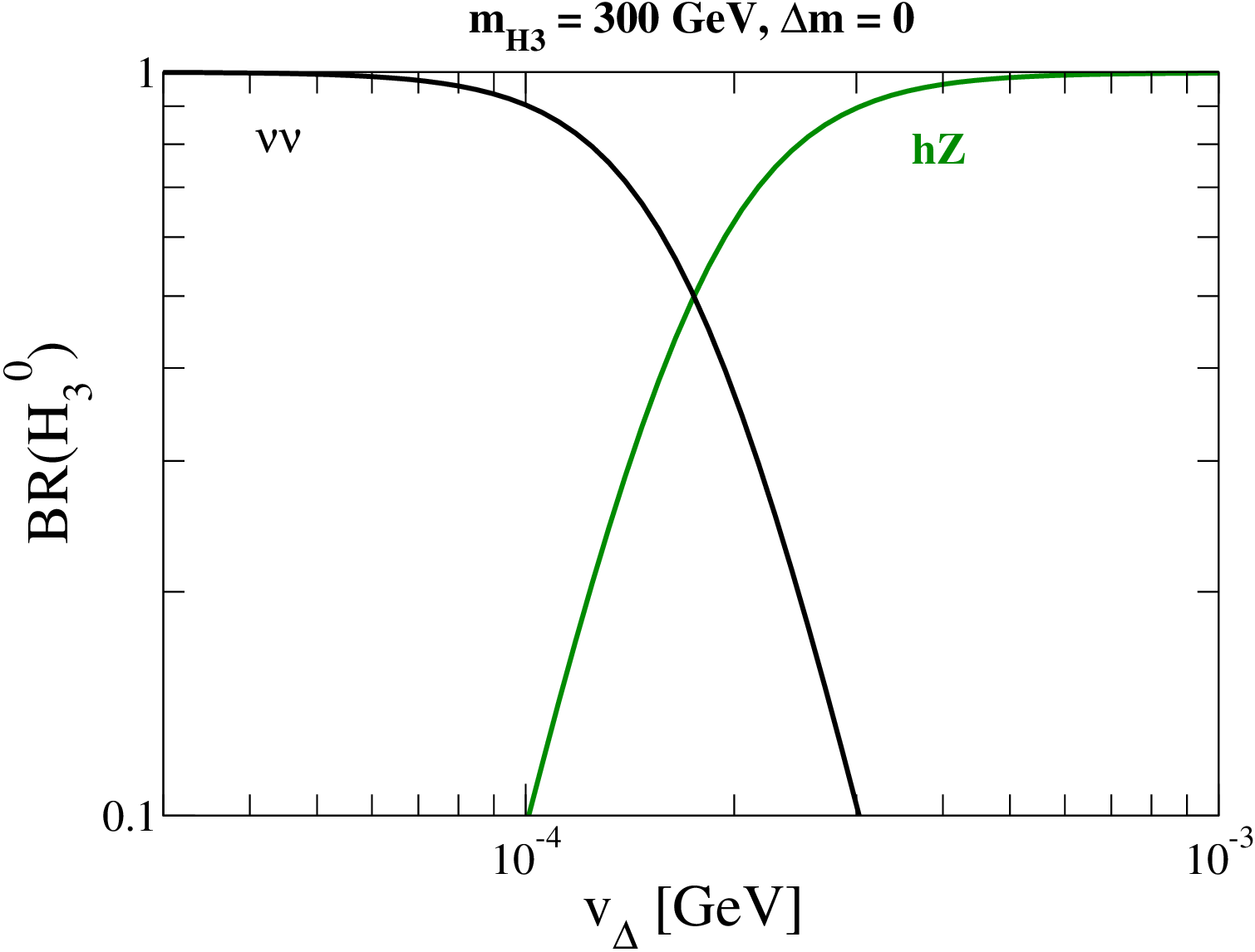}\\\vspace{5mm}
\includegraphics[width=60mm]{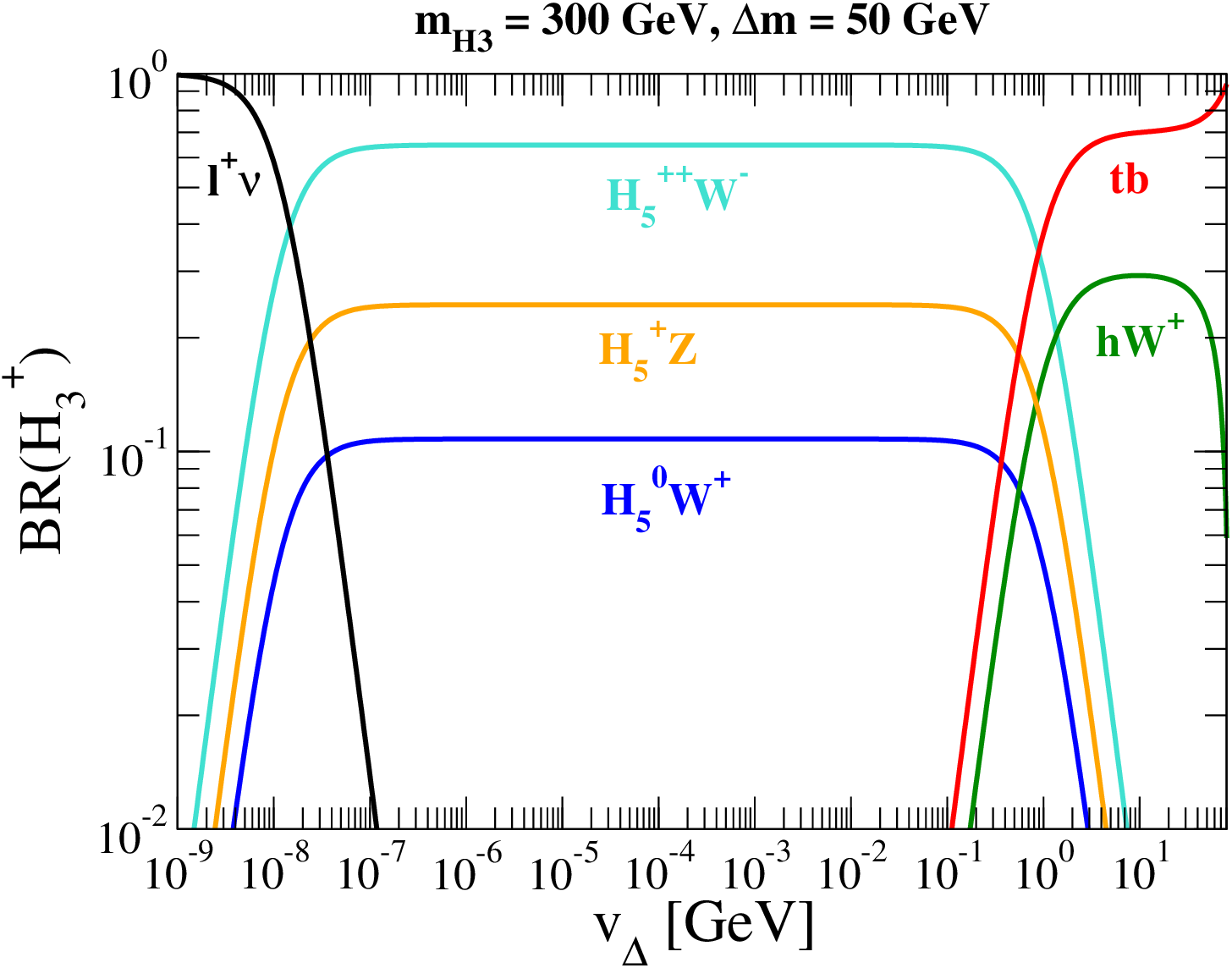}\hspace{10mm}
\includegraphics[width=60mm]{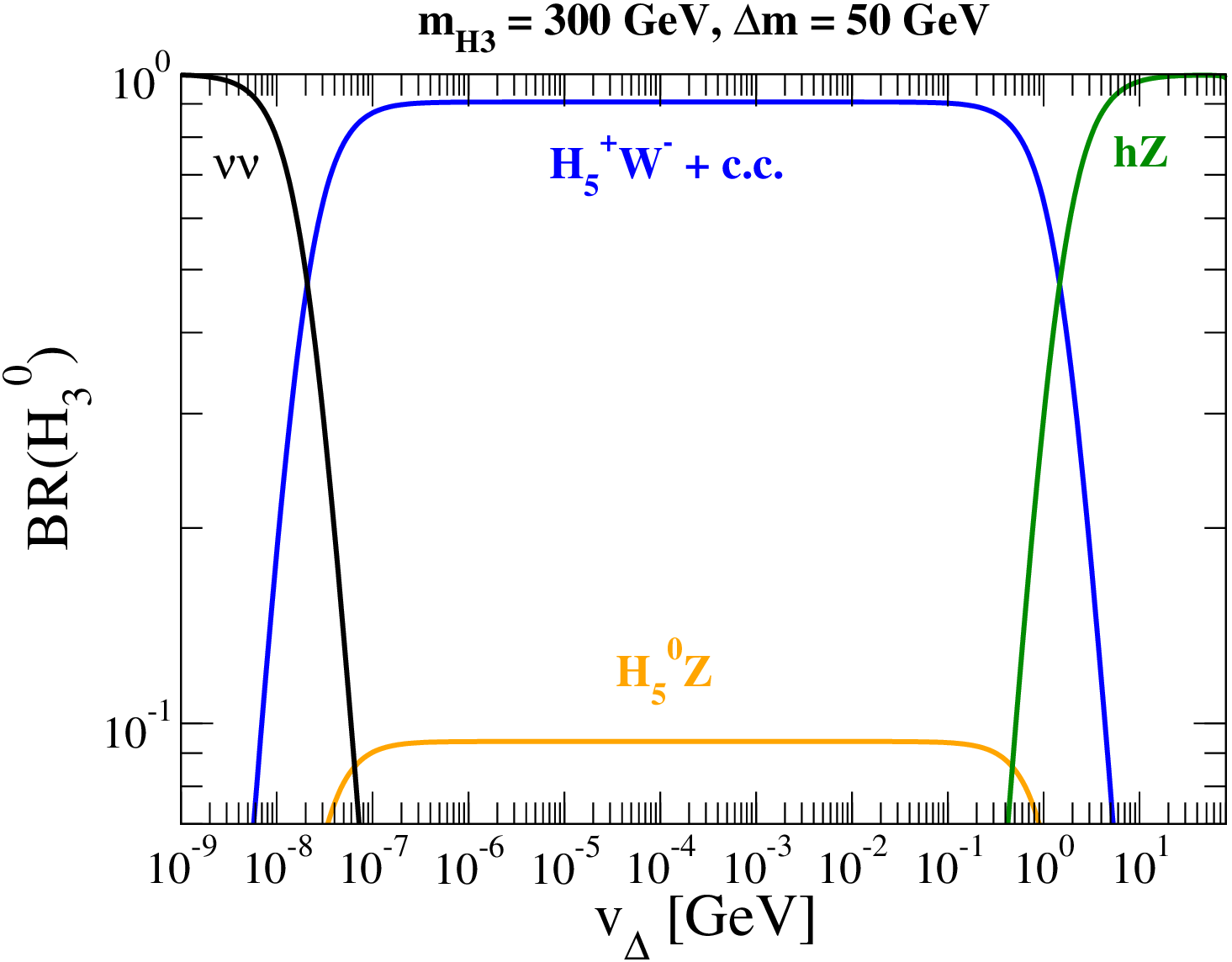}\\\vspace{5mm}
\includegraphics[width=60mm]{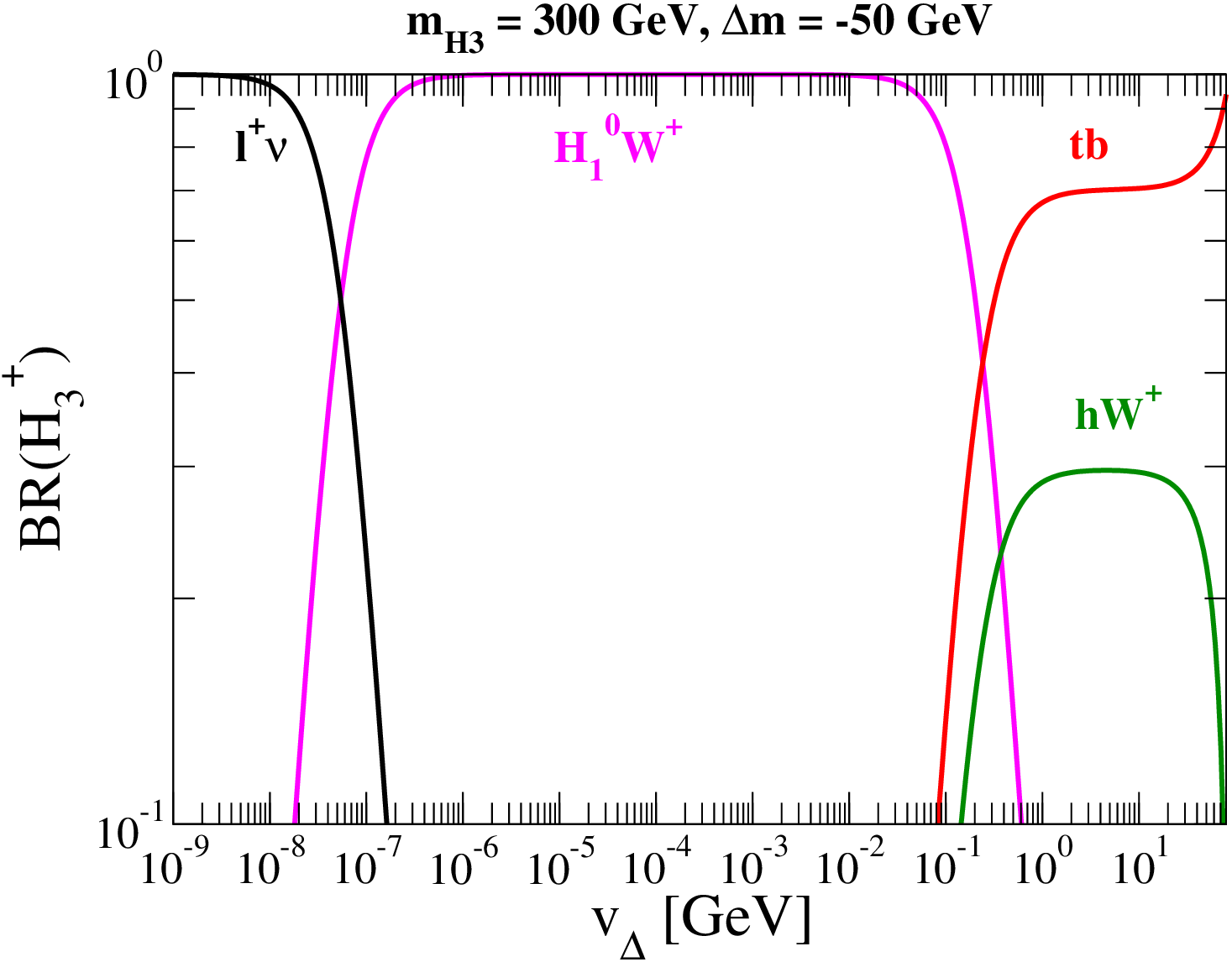}\hspace{10mm}
\includegraphics[width=60mm]{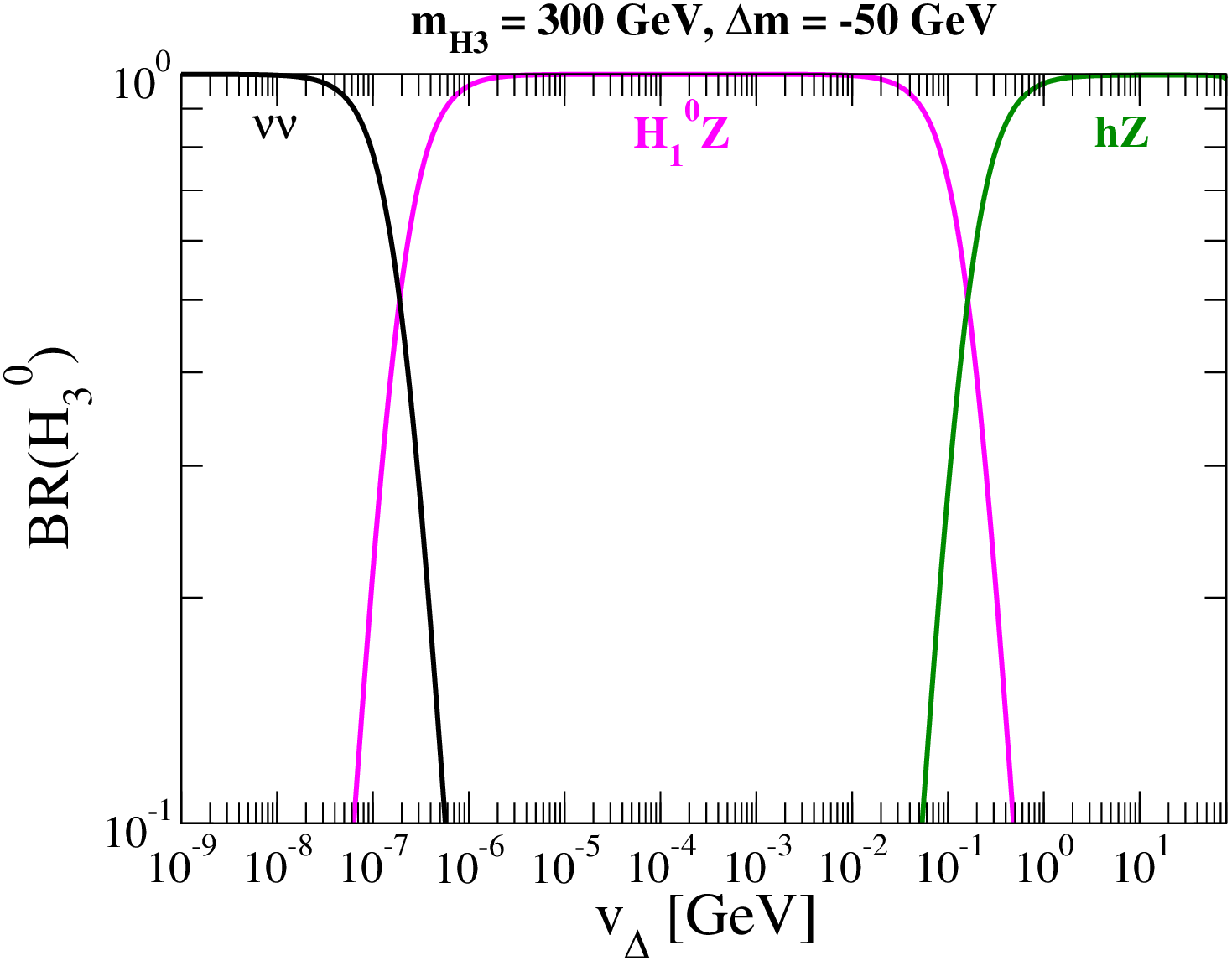}
\end{center}
\caption{Same as Fig.~\ref{decay_150_vd}, but for $m_{H_3}=300$ GeV. }
\label{decay_300_vd}
\end{figure}

In Fig.~\ref{decay_150_vd}, the decay branching ratios of $H_3^+$ and $H_3^0$ are shown as 
a function of $v_\Delta$ for $m_{H_3}=150$ GeV.  The mass difference $\Delta m$ is taken to be 0, 50 GeV and $-50$ GeV in the top, middle and bottom plots, respectively. 
From the top two figures, it is seen that the dominant decay modes of $H_3^+$ ($H_3^0$)  change from $\ell^+\nu$ ($\nu\nu$) to $\tau^+\nu$ or $c\bar{s}$ ($b\bar{b}$) at around $v_\Delta =10^{-3}$ GeV. 
In the case of $\Delta m=50$ GeV (middle plots) and for a wide range $10^{-8}\lesssim v_\Delta \lesssim 10$ GeV, the 3-plet Higgs bosons mainly decay into a 5-plet Higgs boson and a weak gauge boson, {\it i.e.}, $H_3^+ \to H_5^{++} W^-$, $H_3^+ \to H_5^+ Z$ and $H_3^+ \to H_5^0 W^+$ for $H_3^+$ decays and $H_3^0\to H_5^\pm W^\mp$ and $H_3^0\to H_5^0 Z$ for $H_3^0$ decays. 
On the other hand, in the case of $\Delta m=-50$ GeV (bottom plots), the main decay modes of $H_3^+$ ($H_3^0$) are $H_1^0 W^+$ ($H_1^0 Z$) in the range of $10^{-8} \lesssim v_\Delta \lesssim 10$ GeV.

Fig.~\ref{decay_300_vd} shows the decay branching ratios of $H_3^+$ and $H_3^0$ as a function of $v_\Delta$ for $m_{H_3}=300$ GeV.  The mass difference $\Delta m$ is taken to be 0, 50 GeV and $-50$ GeV in the top, middle and bottom plots, respectively. 
When $\Delta m =0$ (top plots), the main decay modes of $H_3^+$ ($H_3^0$) change from $\ell^+\nu$ ($\nu\nu$) 
to $t\bar{b}$ and $hW^+$ ($hZ$) at $v_\Delta \simeq 10^{-4}$ GeV. 
When $\Delta m = 50$ GeV (middle plots) and $\Delta m = -50$ GeV (bottom plots), 
the main decay modes are the same as in the case of $m_{H_3}=150$ GeV in the range of $10^{-7} \lesssim v_\Delta\lesssim 1$ GeV. 

\begin{figure}[t]
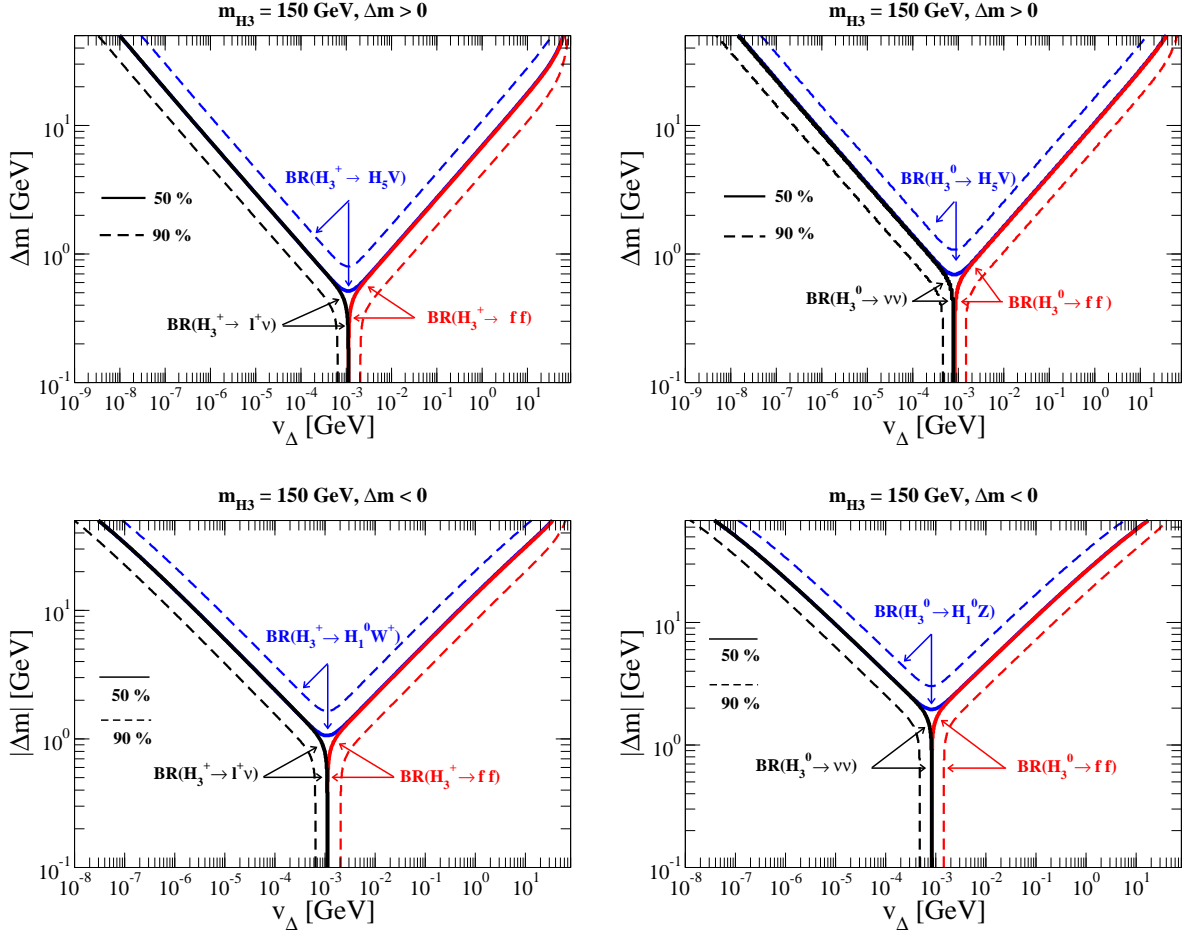

\begin{center}
\includegraphics[width=75mm]{contour_H3p_150_v3.eps}\hspace{5mm}
\includegraphics[width=75mm]{contour_H30_150_v2.eps}\\\vspace{5mm}
\includegraphics[width=75mm]{contour_H3p_150_minus.eps}\hspace{5mm}
\includegraphics[width=75mm]{contour_H30_150_minus.eps}
\caption{
Contour plots of the decay branching ratios of $H_3^{+}$ (left column) and $H_3^0$ (right column) on the $v_\Delta$-$\Delta m$ plane. 
We take $m_{H_3}=150$ GeV, $m_h$=125 GeV and $\alpha=0$ in all the plots. 
The upper (lower) two plots show the case with $\Delta m>0$ ($\Delta m <0$).
Each solid (dashed) curve represents the branching ratio of 50\% (90\%) for the corresponding decay mode indicated by the arrow.}
\label{contour1}
\end{center}
\end{figure}

\begin{figure}[t]
\begin{center}
\includegraphics[width=75mm]{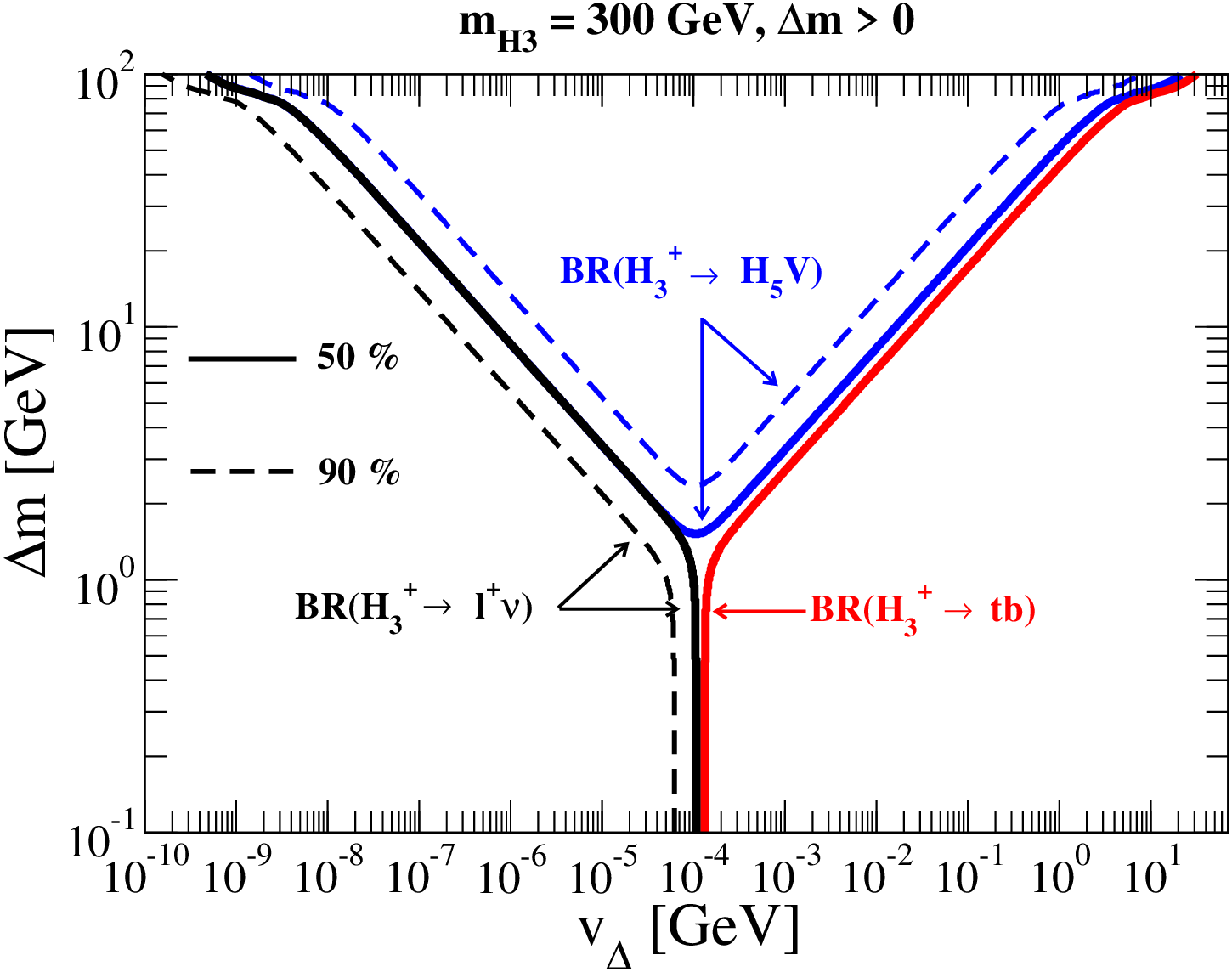}\hspace{5mm}
\includegraphics[width=75mm]{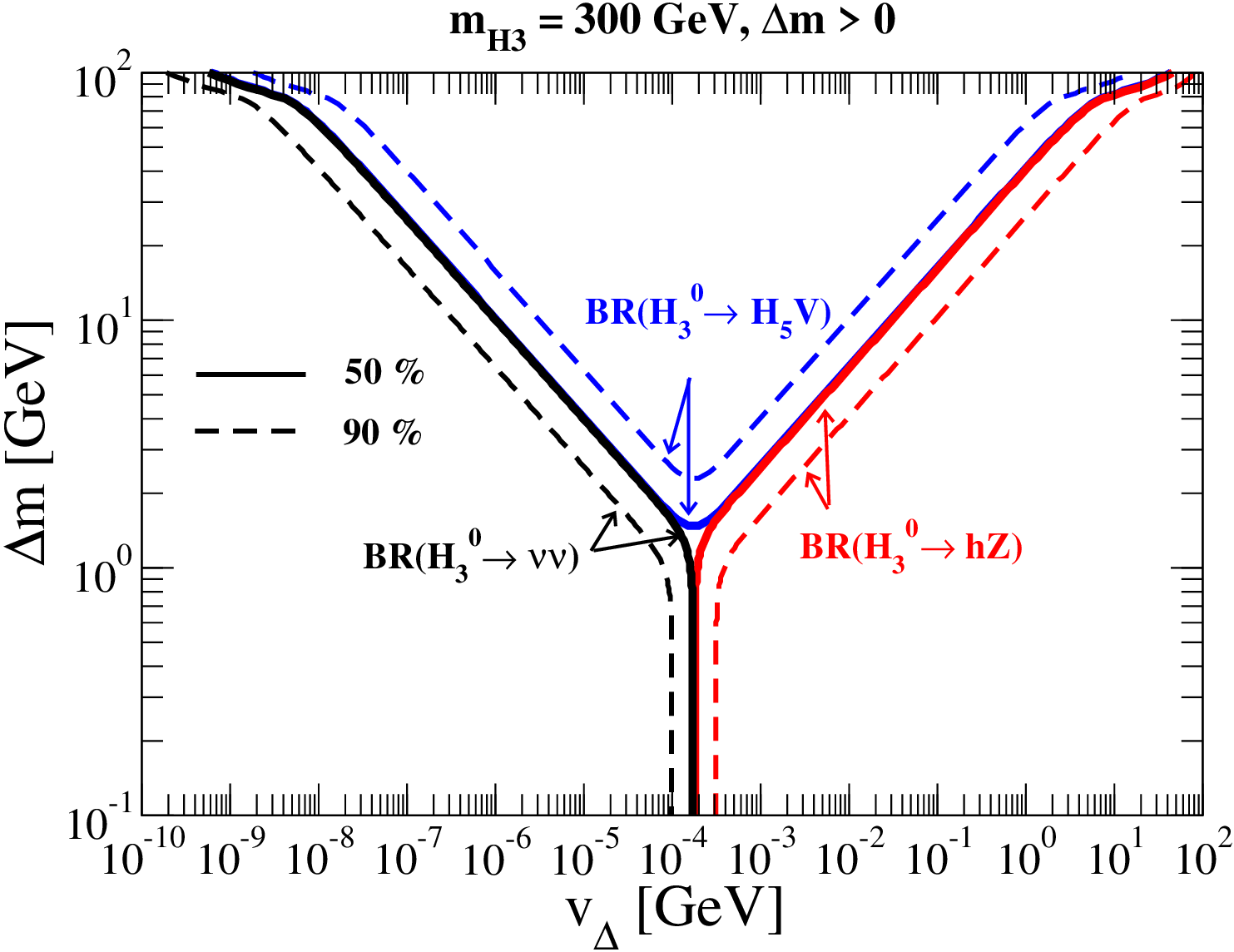}\\
\vspace{5mm}
\includegraphics[width=75mm]{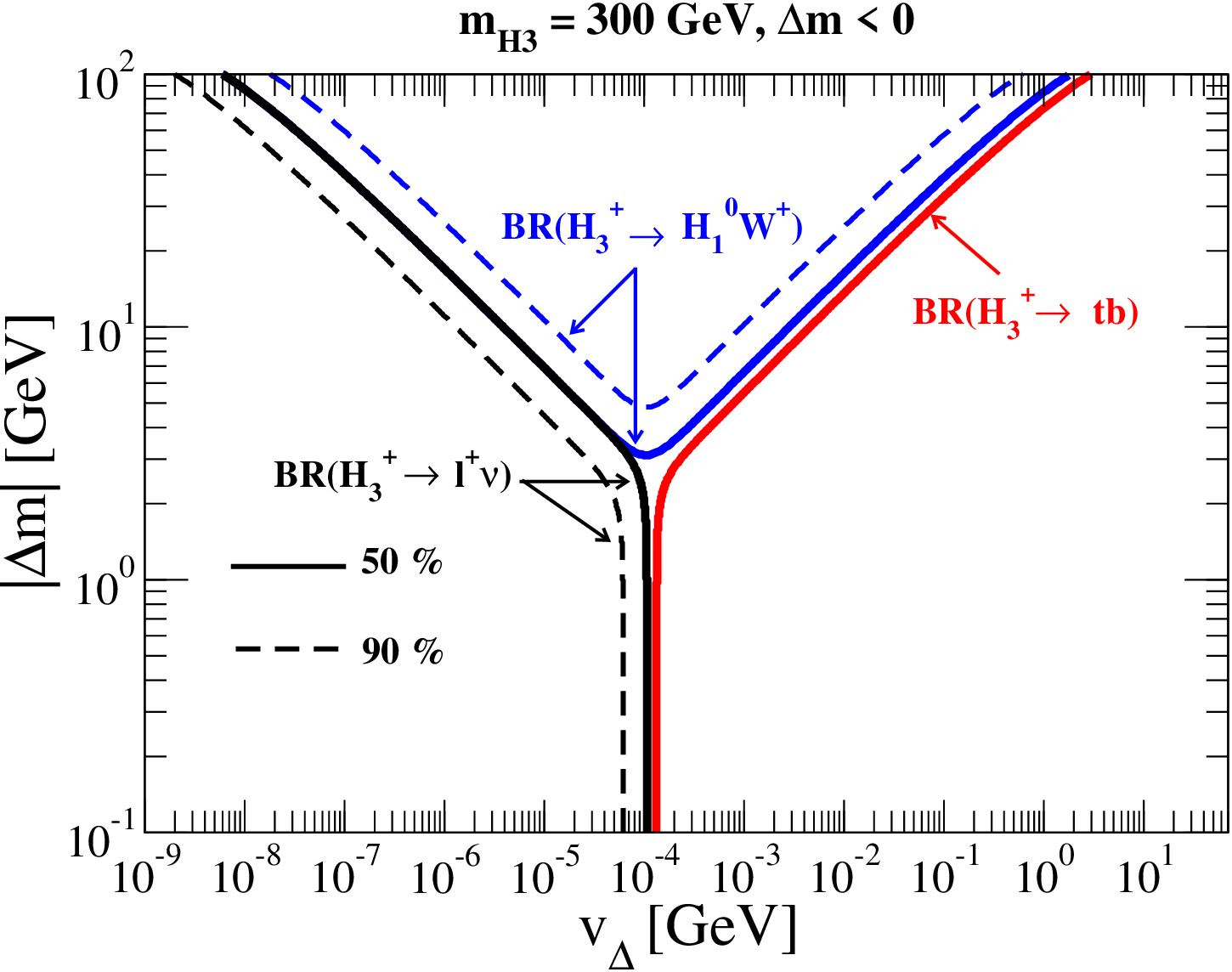}\hspace{5mm}
\includegraphics[width=75mm]{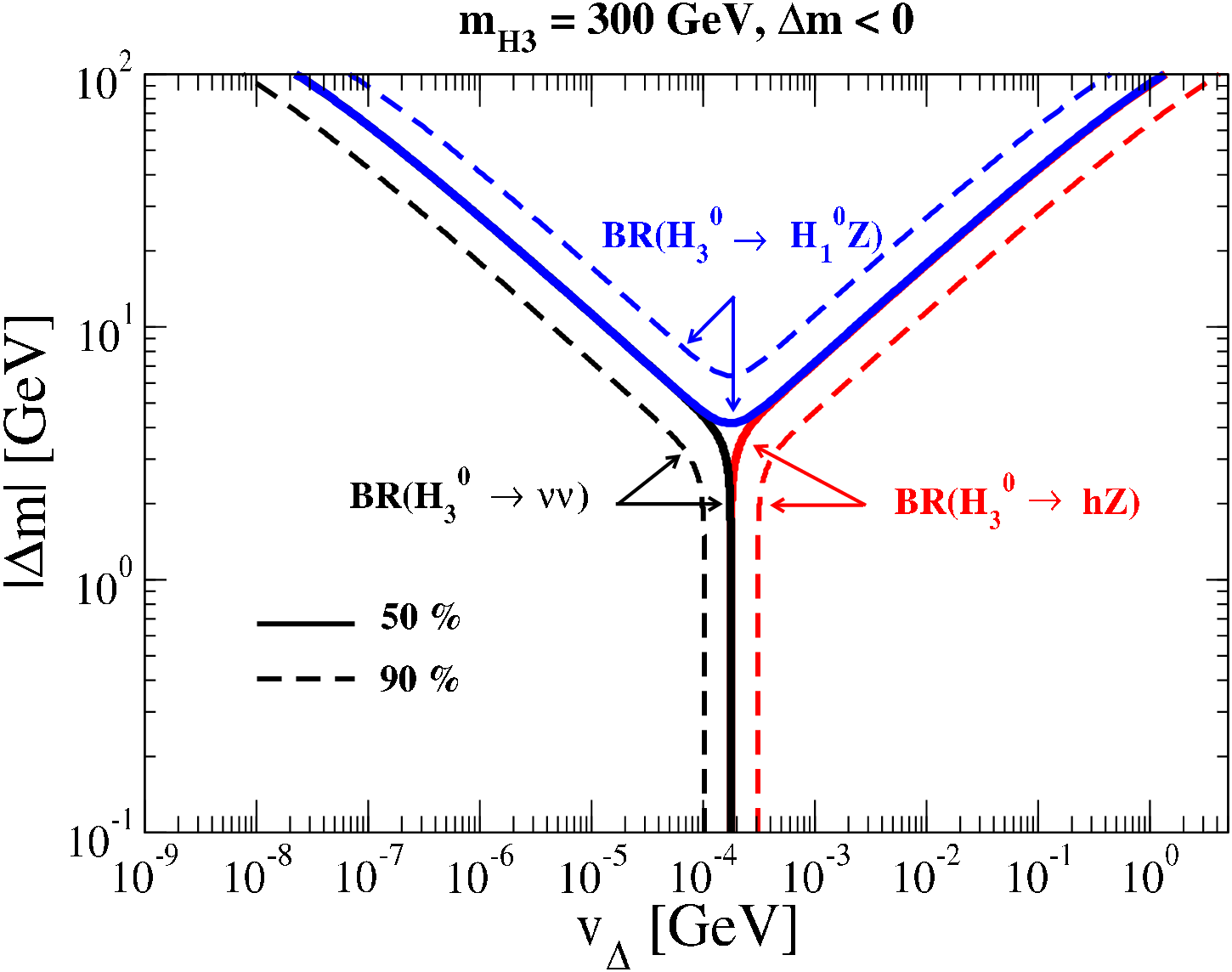}
\caption{Same as Fig.~\ref{contour1}, but for $m_{H_3}=300$ GeV.}
\label{contour2}
\end{center}
\end{figure}

In Fig.~\ref{contour1}, we give the contour plots of the decay branching ratios of $H_3^+$ and $H_3^0$ for $m_{H_3}=150$ GeV.  The mass difference $\Delta m$ is taken to be positive (negative) in the upper (lower) two figures.  
In this figure, BR$(H_3^{+}\to H_5V)$ and BR$(H_3^0\to H_5V)$ denote the sums of the decay branching ratios of the modes with a 5-plet Higgs boson and a gauge boson. 
BR$(H_3^+\to ff)$ and BR$(H_3^0\to ff)$ denote the sum of the decay branching ratios of $H_3^+\to \tau^+\nu$ and $H_3^+\to c\bar{s}$ and that of $H_3^0\to b\bar{b}$ and $H_3^0\to \tau^+\tau^-$, respectively. 
Similar to Fig.~\ref{contour3}, it is seen that there are three distinct regions in this plane. 
In the small $\Delta m$ and small (large) $v_\Delta$ region, the main decay modes are $\ell^+\nu$ ($\tau^+\nu$) for $H_3^+$ and $\nu\nu$ ($b\bar{b}$) for $H_3^0$. 
On the other hand, in the large $\Delta m$ region, the decay modes associated with a 5-plet (singlet) Higgs boson dominate in the case of $\Delta m >0$ ($\Delta m <0$). 
We notice that the regions where the $H_5V$ decay is dominant are wider than the corresponding one where the $H_1^0V$ decay is dominant.  This is because of a larger number of decay modes in $H_5V$.

Fig.~\ref{contour2} shows the contour plots of the branching ratios of the 3-plet Higgs bosons for $m_{H_3}=300$ GeV.  In the plots of the left column, there is no dashed curve corresponding to the branching ratio of 90\% for the $H_3^+\to t\bar{b}$ decay mode. 
This is because the $H_3^+\to hW^+$ decay mode is also kinematically allowed at the same time when the $H_3^+\to t\bar{b}$ is open, and the former amounts to around 30\%. 

We here comment on the decays of the singlet Higgs boson $H_1^0$. 
When we take $\alpha=0$, the decay property of $H_1^0$ is similar to that of $H_5^0$. 
In the case of $\Delta m \leq 0$, $H_1^0$ can decay into $\nu\nu$ ($W^+W^-$ or $ZZ$) for smaller (larger) values of $v_\Delta$. 
When $\Delta m > 0$, $H_1^0$ can decay into a 3-plet Higgs boson and a weak gauge boson. 
If $\alpha\neq 0$, $H_1^0$ can mix with $h$ and can thus decay into fermion pairs via the mixing in addition to the above-mentioned modes. 

\begin{figure}[t]
\begin{center}
\includegraphics[width=100mm]{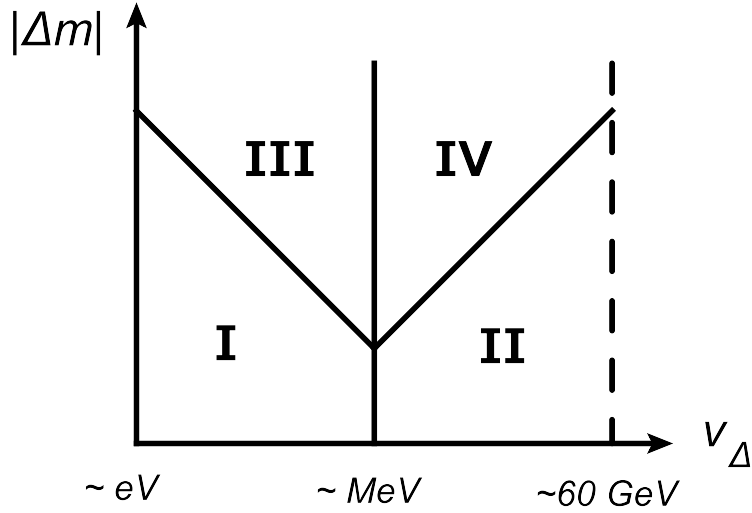}
\end{center}
\caption{Four regions are schematically shown on the $v_\Delta$-$|\Delta m|$ plane.  }
\label{schematic}
\end{figure}

Throughout this section, the decay properties of the 5-plet Higgs bosons and the 3-plet Higgs bosons can be separately considered for four different regions in the $v_\Delta$-$\Delta m$ plane, as schematically shown in Fig.~\ref{schematic}. 
In Region~I, all the triplet-like Higgs bosons mainly decay leptonically: 
\begin{align}
&H_5^{++}\to \ell^+\ell^+,\quad H_5^{+}\to \ell^+\nu,\quad H_5^0\to \nu\nu,\notag\\
&H_3^+\to \ell^+\nu,\quad H_3^0\to \nu\nu.
\end{align}
In this region, the mass of the 5-plet Higgs bosons is constrained to be $m_{H_5}\gtrsim 400$ GeV by the search at the LHC for doubly-charged Higgs bosons decaying into same-sign dileptons~\cite{LHCll}. 
In Region~II, the 5-plet Higgs bosons mainly decay into the weak gauge boson pairs, 
while the 3-plet Higgs bosons decay into the fermion pairs. 
When the mass of the 3-plet Higgs bosons is less than the top quark mass, the main decay modes are 
\begin{align}
&H_5^{++}\to W^+W^+,\quad H_5^{+}\to W^+Z,\quad H_5^0\to W^+W^-/ZZ,\notag\\
&H_3^+\to \tau^+\nu/c\bar{s},\quad H_3^0\to b\bar{b}. \label{S2}
\end{align}
For Region~III and Region~IV, one has to separately consider the cases whether the sign of $\Delta m$ is positive or negative.  In the case of $\Delta m > 0$, the 5-plet Higgs bosons  
mainly decay into the lepton pairs (weak gauge boson pairs) in Region III (Region IV). 
The 3-plet Higgs bosons mainly decay into a 5-plet Higgs boson and a weak gauge boson: 
\begin{align}
&H_5^{++}\to \ell^+\ell^+~(W^+W^+),\quad H_5^{+}\to \ell^+\nu~(W^+Z),\quad H_5^0\to \nu\nu~(W^+W^-/ZZ),\notag\\
&H_3^+\to  H_5^{++} W^{-}/H_5^{+} Z/H_5^0 W^+,\quad 
H_3^0\to H_5^\pm W^{\mp }/H_5^0Z.
\end{align}
In the case of $\Delta m<0$, 
the main decay modes in both Region~III and Region~IV are
\begin{align}
&H_5^{++}\to H_3^+ W^+,\quad H_5^{+}\to H_3^+ Z/H_3^0 W^+,\quad H_5^0\to H_3^\pm W^\mp/H_3^0 Z\notag\\
&H_3^+\to H_1^0W^+ ,\quad H_3^0\to H_1^0Z.
\end{align}

\section{Phenomenology at the LHC \label{sec:pheno}}

In this section, we discuss how the custodial symmetry of the GM model can be tested at the LHC.  There are characteristic features of the triplet-like Higgs bosons that mainly originate from the triplet field $\Delta$ of the model. 
(1) The masses of the Higgs bosons belonging to the same $SU(2)_V$ multiplet are the same. 
(2) The 5-plet and the singlet Higgs bosons have the Gauge-Gauge-Scalar type of couplings as listed in Table~\ref{GGS}, but not the Yukawa couplings given in Eq.~(\ref{Yukawa1}), while
the 3-plet Higgs bosons have the Yukawa couplings, but not the Gauge-Gauge-Scalar type of couplings.  
These features can be used to test the custodial symmetry of the GM model. 

As is discussed in the previous section, it is important to study the decay pattern of the triplet-like Higgs bosons.  In particular, feature (2) mentioned above can be most clearly tested in Region~II because the 5-plet (3-plet) Higgs bosons mainly decay into weak gauge boson pairs (fermion pairs) in this region.  In the following discussion, we focus on Region~II and the detectability of the 5-plet and 3-plet Higgs bosons.

\subsection{Production modes}

\begin{figure}[!t]
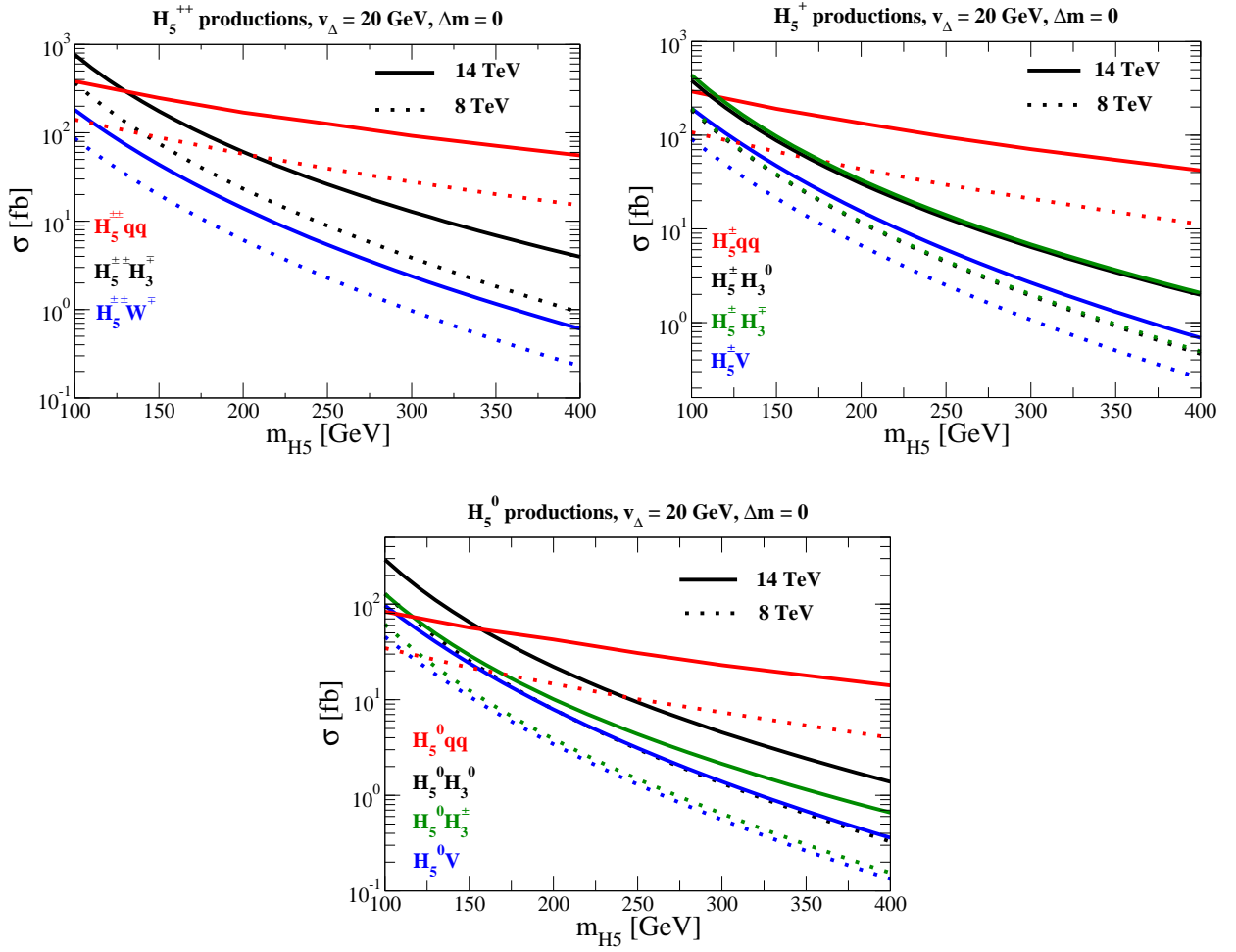

\begin{center}
\includegraphics[width=80mm]{cs_H5pp.eps}\hspace{3mm}
\includegraphics[width=80mm]{cs_H5p.eps}\\\vspace{5mm}
\includegraphics[width=80mm]{cs_H50.eps}
\end{center}
\caption{Production cross sections of the 5-plet Higgs bosons in various processes as a function of $m_{H_5}$. 
The upper-left, upper-right, and bottom plots show the production cross sections for $H_5^{\pm\pm}$, $H_5^{\pm}$, and $H_5^0$, respectively.  In all the plots, we take $v_\Delta =20$ GeV and $\Delta m =0$. 
The LHC collision energy is assumed to be 8 TeV (dashed curves) and 14 TeV (solid curves). }
\label{cross}
\end{figure}

There are several production modes for the 5-plet Higgs bosons $H_5$ and the 3-plet Higgs bosons $H_3$, as listed below. 
Throughout this section, $q,q',Q,Q'$ and those with bars denote light quarks and anti-quarks.

\begin{description}
\item[1.] {\bf The Drell-Yan process}\\
$H_5$ and $H_3$ can be produced in pairs via $\gamma$ and $Z$, {\it e.g.}, $pp\to H_5H_5$ and $pp\to H_3H_3$.  The cross section is determined by the gauge coupling as well as the Higgs masses $m_{H_5}$ and $m_{H_3}$, independent of the value of $v_\Delta$.  
\item[2.] {\bf The mixed Drell-Yan (mDY) process}\\
$H_5$ and $H_3$ can be produced at the same time, {\it e.g.}, $pp\to H_5H_3$, which we call the mixed Drell-Yan (mDY) process to be separated from the usual Drell-Yan process mentioned above.  The cross section is proportional to $c_H^2$, and is thus relatively suppressed in comparison with the Drell-Yan process, especially in the large $v_\Delta$ case. 
\item[3.] {\bf The weak vector boson fusion (VBF) process}\\
The single production of $H_5$ occurs via the $qQ\to H_5$ process. 
The cross section is proportional to $v_\Delta^2$, so that this mode can be important in the large $v_\Delta$ case. 
\item[4.] {\bf The weak vector boson associated process}\\
In addition to the VBF process, $H_5$ can also be produced in association with a weak gauge boson, {\it e.g.}, $q\bar{q}'\to H_5V$.  The cross sections of such modes are proportional to $v_\Delta^2$ as for the VBF production mode.  Thus, this mode can also become important when the VBF process is important. 
\item[5.] {\bf The Yukawa process}\\
$H_3$ can be produced through the Yukawa interactions given in Eq.~(\ref{Yukawa1}) as the gluon fusion process for the SM Higgs boson: $gg\to H_3^0$.  
There are t-channel $H_3^\pm$ and $H_3^0$ production modes: $gb\to tH_3^-$ and $gb\to bH_3^0$. 
These production cross sections are proportional to $\tan^2\theta_H$. 
\item[6.] {\bf The top quark decay}\\
When $m_{H_3}$ is smaller than the top quark mass, $H_3^\pm$ can be produced from 
the top quark decay. The decay rate of the $t\to bH_3^{\pm}$ depends on $\tan^2\theta_H$. 
\end{description}

Among these production processes, channels 3 and 4 can be useful to discriminate the GM model from the others with doubly-charged Higgs bosons 
and to test the mass degeneracy of $H_5$. 
In the HTM, for example, the doubly-charged Higgs boson can in principle be produced via the VBF and the vector boson associated processes. 
However, these cross sections are much suppressed due to the tiny triplet VEV required by the electroweak rho parameter. 
In the GM model, $v_\Delta$ can be of order 10 GeV, so that these production processes become 
useful. 
The mDY process is also a unique feature of the GM model because the Higgs bosons $H_5$ and $H_3$ having different decay properties are produced at the same time. 
In particular, when Region~II is realized, the main decay modes of these two Higgs bosons are distinctly different.  Thus, this process can be useful not only to test the mass degeneracy of $H_3$ but also to distinguish the model from the others also having doubly-charged and/or singly-charged Higgs bosons. 

In Fig.~\ref{cross}, production cross sections of the 5-plet Higgs bosons from channels 2, 3, and 4 are shown as a function of $m_{H_5}$ for the LHC running at 8 and 14 TeV.  We take $v_\Delta=20$ GeV and $\Delta m = 0$ as an example in all the plots.  It is noted that the dominant production mechanism is the VBF process for a sufficiently large $m_{H_5}$.

\subsection{Signal and background analysis}

We will first discuss the VBF process and the vector boson associated process to study 
the mass degeneracy among $H_5^{\pm\pm}$, $H_5^\pm$ and $H_5^0$.  
Then we turn to the mDY process. 

Let us consider the case with $m_{H_3}=150$ GeV, $\Delta m = 10$ GeV ({\it i.e.}, $m_{H_5}=140$ GeV) and $v_\Delta = 20$ GeV as an example in Region~II. 
In this case, the 5-plet Higgs bosons decay into gauge boson pairs almost 100\% 
(the branching fractions of $H_5^0\to W^+W^-$ and $H_5^0\to ZZ$ being 67\% and 33\%, respectively).  On the other hand, $H_3^\pm$ decays to $\tau^\pm\nu$ at 66\% and $cs$ at 29\%, and $H_3^0$ decays to $b\bar{b}$ at 89\%. 
We note that the branching fraction of $t\to H_3^+ b$ here is around 0.4\%. 
The upper limit of the top quark decay into a charged Higgs boson and the bottom quark is 2-3\% 
in the case where the charged Higgs boson mass is between 80 and 160 GeV, 
under the assumption that the charged Higgs boson decays to $\tau\nu$ at 100\%
~\cite{top_decay}.  Thus, the selected parameter set is allowed by the constraint from the top quark decays. 
The signal events from the VBF production processes for the 5-plet Higgs bosons are given by
\begin{align}
qQ &\to H_5^{\pm\pm} q'Q'\to W^\pm W^\pm jj,\notag\\
qQ &\to H_5^\pm q'Q'\to W^\pm Z jj,\notag\\
qQ &\to H_5^0 q'Q'\to W^\pm W^\mp jj/ZZ jj. 
\end{align}
From the vector boson associated processes, we have the following events
\begin{align}
q\bar{q}' &\to H_5^{\pm\pm} W^\mp \to W^\pm W^\pm jj\notag\\
q\bar{q} &\to H_5^\pm W^\mp \to W^\pm Zjj,\quad q\bar{q}' \to H_5^\pm Z \to W^\pm Zjj,\notag\\
q\bar{q} & \to H_5^0 Z \to W^+W^-jj/ZZjj,\quad q\bar{q}' \to H_5^0 W^\pm \to W^+W^-jj/ZZjj,  
\end{align}
where the associated weak gauge bosons are assumed to decay hadronically so that they have the same final states as the VBF process. 
Moreover, we consider the case where the weak gauge bosons produced from the decay of $H_5$ decay leptonically.  Then the final states of the signal events have same-sign (SS) dileptons plus dijets and missing transverse energy ($\ell^\pm\ell^\pm jj \slashed{E}_T$) for the $H_5^{\pm\pm}$ production mode, where $\ell^\pm$ denotes collectively the light leptons $e^\pm$ and $\mu^\pm$ hereafter.
The final state of the $H_5^{\pm}$ production mode includes trileptons plus dijets and missing transverse energy ($\ell^\pm\ell^\pm \ell^\mp jj \slashed{E}_T$), while that for the $H_5^0$ production mode has opposite-sign (OS) dileptons plus dijets and missing transverse energy 
($\ell^\pm\ell^\mp jj \slashed{E}_T$).  
The corresponding background events for these signal events are from the 
$W^\pm W^\pm jj$ for the $H_5^{\pm\pm}$ production, 
$W^\pm Z jj$ for the $H_5^{\pm}$ production, and 
$t\bar{t}$, $W^\pm W^\mp jj$ and $ZZjj$ for the $H_5^{0}$ production. 

We simulate the signal and the background event rates by using MadGraph~5~\cite{MG5} at the parton level for the cases where the LHC operates at the center-of-mass (CM) energy $\sqrt{s}$ of 8 TeV and 14 TeV. 
We impose the following basic kinematic cuts
\begin{align}
&p_T^j > 20~\text{GeV},\quad p_T^\ell > 10~\text{GeV},\quad|\eta^j| < 5,\quad |\eta^\ell| < 2.5,\quad \Delta R^{jj} > 0.4, \label{basic}
\end{align}
where $p_T^j$ and $p_T^\ell$ are the transverse momenta of the jet and the lepton, respectively, $\eta^j$ and $\eta^\ell$ are the pseudorapidities of the jet and the lepton, respectively, and $\Delta R^{jj}$ is the distance between the two jets. 
The cross sections for the signal and background events are listed in Table~\ref{sig_and_bg},
where the signal cross section includes contributions from the VBF production and the vector boson associated production.  An integrated luminosity of 100 fb$^{-1}$ is assumed in the simulations.
In this table, the signal significance is defined by 
\begin{align}
\mathcal{S} = S/\sqrt{S+B}, \label{significance}
\end{align}
where $S$ and $B$ are the numbers of the signal and background events, respectively.
The significance of the $\ell^\pm\ell^\pm jj \slashed{E}_T$ event from the $H_5^{\pm\pm}$ production process exceeds 5 even using simply the basic cuts. 
However, the significances for the remaining two events from the $H_5^\pm$ and $H_5^0$ 
production processes are less than 1. 
For the $\ell^\pm\ell^\mp jj \slashed{E}_T$ event, in particular,
the background is larger than the signal by 3 to 4 orders of the magnitude because of the 
huge $t\bar{t}$ background.

\begin{table}[t]
\begin{center}
{\renewcommand\arraystretch{1.2}
\begin{tabular}{|c||c|c|c||c|c|c||c|c|c|}\hline
&\multicolumn{3}{c||}{$\ell^\pm\ell^\pm jj \slashed{E}_T$}&\multicolumn{3}{c||}
{$\ell^\pm\ell^\pm\ell^\mp jj \slashed{E}_T$}&\multicolumn{3}{c|}
{$\ell^\pm\ell^\mp jj \slashed{E}_T$}\\\hline
Cuts&$H_5^{\pm\pm}jj$&$W^\pm W^\pm jj$&$\mathcal{S}$&
$H_5^{\pm}jj$&$W^\pm Z jj$&$\mathcal{S}$&
$H_5^{0}jj$& $t\bar{t}/VVjj$ &$\mathcal{S}$  \\\hline\hline
Basic&3.71&3.48 &13.8&0.61&45.9&0.89&1.15&4.39$\times 10^{3}$& 0.17\\
&(8.72)&(8.13) &(21.2)&(1.60)&(1.39$\times 10^2$)&(1.35)&(2.76)&(1.77$\times 10^{4}$)& (0.21)\\\hline
$\Delta \eta^{jj}$&1.82 & 0.20 & 12.8 & 0.33 &4.42 & 1.51 & 0.51 & 30.7 & 0.91
\\
&(5.68) & (0.65) & (22.6) & (0.98) &(15.6) &(2.41) & (1.42) & (1.99$\times 10^2$) & (1.00)
\\\hline
$M_T$&1.80 & 0.05 & 13.2 & 0.33 & 0.07 & 5.22 & 0.48 & 11.4 & 1.39
\\
&(5.58) & (0.12) & (23.4) & (0.98) & (0.46) & (8.17) & (1.36) & (67.4) & (1.64)
\\\hline
$b$-jet veto&-&-&-&-&-&-&0.48&1.82&3.16
\\
&-&-&-&-&-&-&(1.36)&(10.8)&(3.90)
\\\hline
\end{tabular}}
\caption{Signal and background cross sections in units of fb after each kinematic cut, along with the significance $\mathcal{S}$ defined by Eq.~(\ref{significance}) based on an integrated luminosity of 100 fb$^{-1}$.  The numbers without (with) parentheses correspond to the case with a CM energy of $8$ TeV (14 TeV). 
The signal cross section includes contributions from both the VBF production and the vactor boson associated production processes. 
For the $\ell^\pm\ell^\mp jj \slashed{E}_T$ events, we further impose the requirement of the 
$b$-jet veto for each jet to reduce the background, where the $b$-tagging efficiency is take to be 0.6~\cite{ATLAS_bjet}. }
\label{sig_and_bg}
\end{center}
\end{table}

\begin{figure}[!t]
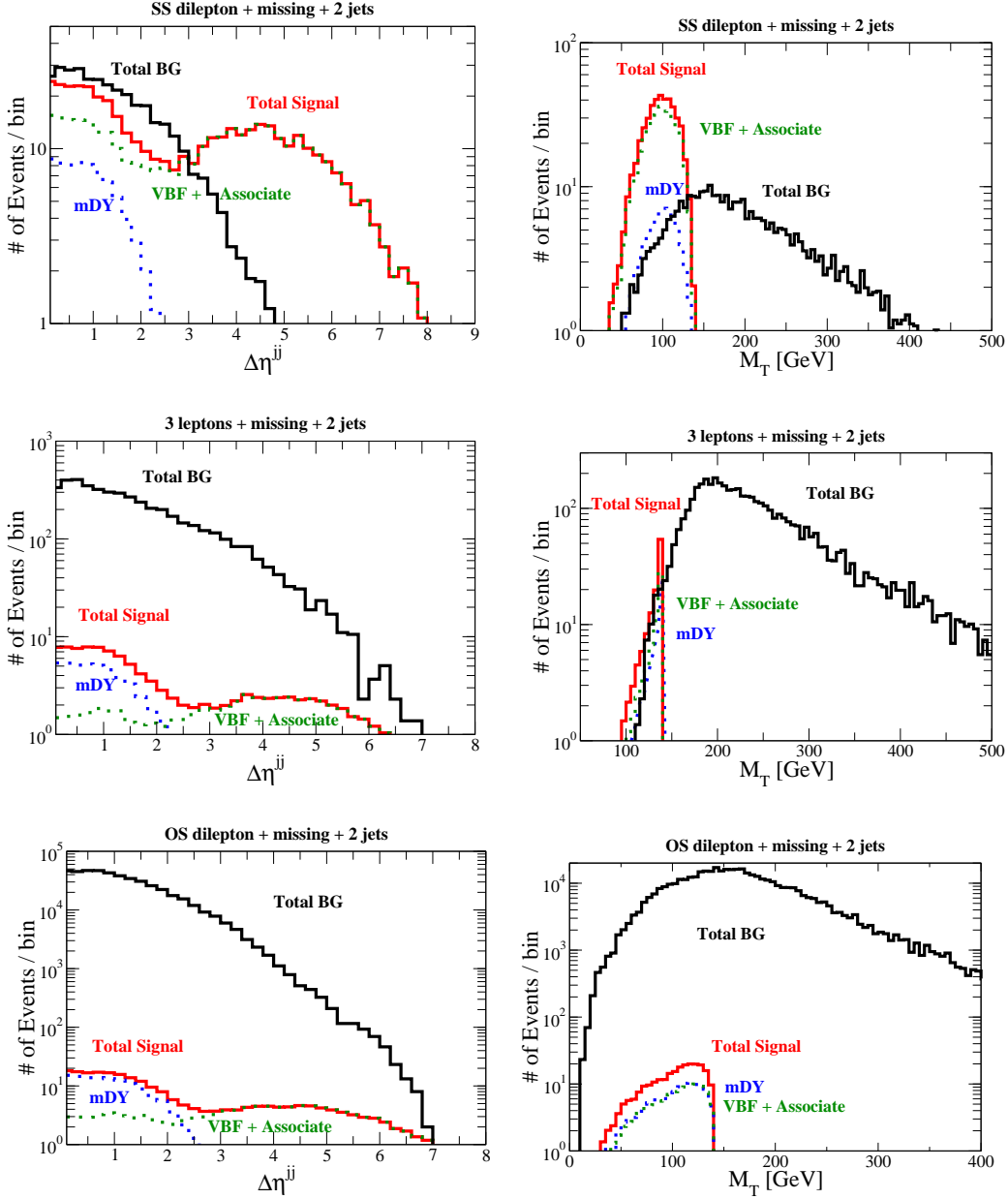

\begin{center}
\includegraphics[width=65mm]{Deleta_lplp.eps}\hspace{6mm}
\includegraphics[width=65mm]{mt_lplp.eps}\\\vspace{5mm}
\includegraphics[width=65mm]{Deleta_lll.eps}\hspace{6mm}
\includegraphics[width=65mm]{mt_lll.eps}\\\vspace{5mm}
\includegraphics[width=65mm]{Deleta_lplm.eps}\hspace{3mm}
\includegraphics[width=65mm]{mt_lplm.eps}
\end{center}
\caption{$\Delta\eta^{jj}$ and $M_T$ distributions for the signal and background events. 
The top, middle and bottom plots show these distributions for the $\ell^\pm\ell^\pm jj \slashed{E}_T$, 
$\ell^\pm\ell^\pm \ell^\mp jj \slashed{E}_T$ and $\ell^\pm\ell^\mp jj \slashed{E}_T$ events, respectively. 
The distributions for the signal events are divided into those from the VBF process and vector boson associated process (green dashed curve), the mDY process (blue dashed curve) and the sum of them (red solid curve). 
The bin size for the $\Delta\eta_{jj}$ ($M_T$) distribution is taken to be 0.2 (5 GeV). 
The integrated luminosity and the CM energy are assumed to be 100 fb$^{-1}$ and 8 TeV, respectively. }
\label{distr}
\end{figure}

To improve the significance, we need to impose additional kinematic cuts. 
Fig.~\ref{distr} shows the distributions of the pseudorapidity gap $\Delta\eta^{jj}$ for the dijet system and the transverse mass~\cite{transverse_mass} in the leptons plus missing transverse energy system for $\sqrt{s}=8$ TeV and the 
integrated luminosity of 100 fb$^{-1}$. 
Explicitly, these two kinematical quantities are defined by 
\begin{align}
M_T^2 &\equiv\left[\sqrt{
M_{\text{vis}}^2+ ({\bm p}_T^{\text{vis}})^2}+| \slashed{\bm p}_T |\right]^2-
\left[{\bm p}_T^{\text{vis}}+ \slashed{\bm p}_T \right]^2, \\
\Delta\eta^{jj}&\equiv |\eta^{j_1}-\eta^{j_2}|, 
\end{align}
where $M_{\text{vis}}$ and ${\bm p}_T^{\text{vis}}$ are the invariant mass and the vector sum of the transverse momenta of the charged leptons, 
respectively, and $\slashed{\bm p}_T$ is the missing transverse momentum determined by the negative sum of visible momenta in the transverse direction.
In Fig.~\ref{distr}, the distributions of $\Delta\eta^{jj}$ and $M_T$ for the signal events from the VBF process plus vector boson associated process and the mDY process are separately indicated by dotted lines. 
The latter production mode will be discussed in details later. 
A significant feature of the VBF process is that the two external quark jets are almost along the beam direction and carry most of the energy of the collider protons.  Therefore, they are mostly detected in the forward regions.
This is seen in the $\Delta\eta^{jj}$ distribution of Fig.~\ref{distr}. 
The end point in the $M_T$ distribution of signals rests at around 140 GeV, corresponding to the mass of the 5-plet Higgs boson. 

According to the above-mentioned observations, we find the following additional kinematic cuts useful in further reducing the backgrounds: 
\begin{align}
\Delta \eta^{jj} >3.5~~(>4.0~~\text{for}~~\ell^\pm\ell^\mp jj \slashed{E}_{T}),\quad 50<M_T <150\text{ GeV}.
\label{cut_add}
\end{align}
The cross sections for the signals and backgrounds in each step of the kinematic cuts are listed in Table~\ref{sig_and_bg}. 
After making the first two cuts, the significances of the events from $H_5^{\pm\pm}$ and $H_5^{\pm}$ achieve $13.2$ and $5.2$, respectively. 
However, the significance of the events from $H_5^0$ is around $1.4$. 
We further require that events with at least one $b$-jet are tagged and rejected in order to reduce the $t\bar{t}$ 
background. The $b$-tagging efficiency is taken to be 0.6~\cite{ATLAS_bjet}. 
By using this cut, the $t\bar{t}$ background events with the final state of $b\bar{b}\ell^+\ell^- \slashed{E}_T$ 
can be reduced to be 16\%. 
Consequently, the signal significance for the $\ell^+\ell^-jj \slashed{E}_T$ event
can reach $3.16$ ($3.90$) with $\sqrt{s}=8$ TeV (14 TeV) after all the cuts discussed above are imposed. 

Next, we focus on the mDY production mode discussed in the previous subsection. 
In order to reconstruct the masses of $H_3$ Higgs bosons, we consider their hadronic decays, namely $H_3^\pm\to cs$ and $H_3^0\to b\bar{b}$. 
The signal events are
\begin{align}
&pp\to H_5^{\pm\pm}H_3^{\mp}\to W^\pm W^\pm cs,\notag\\
&pp\to H_5^{\pm}H_3^{\mp}\to W^\pm Z cs,\quad
pp\to H_5^\pm H_3^{0}\to W^\pm Z b\bar{b},\notag\\
&pp\to H_5^0H_3^\pm \to W^+W^-cs/ZZcs,\quad 
pp\to H_5^0H_3^0\to W^+W^-b\bar{b}/ZZb\bar{b}, \label{sig_process}
\end{align}
where leptonic decays of the weak gauge bosons from the $H_5$ decays are also assumed in this analysis. 
Thus, the final states of the signal events from the mDY process are the same as those from the VBF process 
as well as the associated process. 
Its difference from the VBF process is observed in the $\Delta\eta^{jj}$ distribution of the dijet system. 
In the mDY process, the dijets in the final state come from the decay of the 3-plet Higgs boson, not the external quark jets.  
According to the plots in the left column of Fig.~\ref{distr}, the events from the mDY process concentrates in the $\Delta\eta^{jj}\lesssim 2.5$ region for all the three cases. 
On the other hand, the $M_T$ distributions from the mDY process and the VBF plus associated process are almost the same.  This is because the leptons plus missing transverse energy system come from the decays of $H_5$ in both processes. 
Therefore, we apply the same $M_T$ cut given in Eq.~(\ref{cut_add}) to this analysis, but not the $\Delta\eta^{jj}$ cut. 
In the analysis of the mDY process, the $\ell^\pm \ell^\mp jj \slashed{E}_{T}$ signal events are overwhelmed by the huge background from the $t\bar{t}$ production. 

Table~\ref{sig_and_bg2} lists the cross sections of the signal and the background events after imposing the basic cut and $M_T$ cut.  In addition, the signal significance is given by assuming an integrated luminosity of 100 fb$^{-1}$.  We find that the significances exceed 5 in both cases after imposing the $M_T$ cut. 

\begin{figure}[!t]
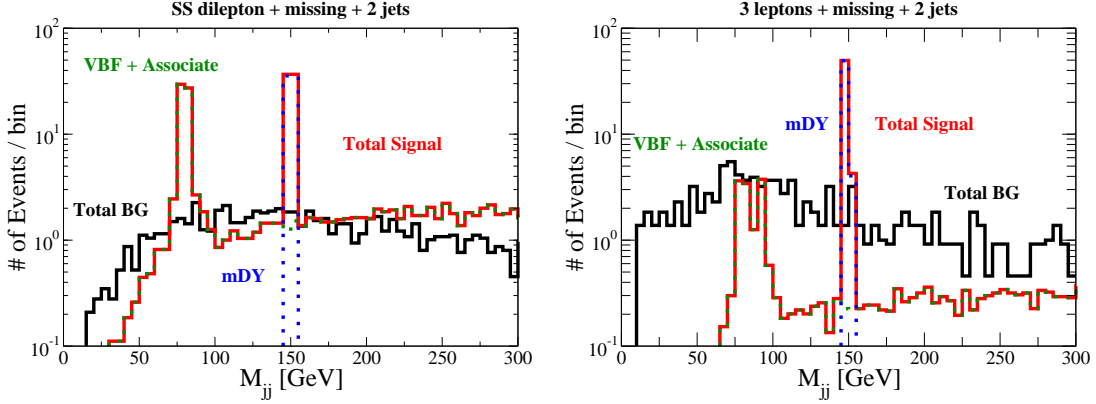

\begin{center}
\includegraphics[width=70mm]{Mjj_lplp.eps}\hspace{3mm}
\includegraphics[width=70mm]{Mjj_llln.eps}
\end{center}
\caption{Invariant mass distribution for the dijets system.  The bin size in this distribution is 5 GeV. 
The distributions for the signal events are divided into those from the VBF process and vector boson associated process (green dashed curve), the mDY process (blue dashed curve) and the sum of them (red solid curve). 
The integrated luminosity and the CM energy are assumed to be 100 fb$^{-1}$ and 8 TeV, respectively. }
\label{distr2}
\end{figure}

Fig.~\ref{distr2} show the dijet invariant mass $M_{jj}$ distributions of the signal and the background events.  These distributions are plotted after imposing the $M_T$ cut.  We can see a peak at around 150 GeV, corresponding to the mass of the 3-plet Higgs bosons, in both the $\ell^\pm \ell^\pm jj \slashed{E}_T$ and $\ell^\pm\ell^\pm \ell^\mp jj \slashed{E}_T$ events. 
This suggests that the mass degeneracy between $H_3^\pm$ and $H_3^0$ can be readily established from the mDY process. 
First, the $\ell^\pm \ell^\pm jj \slashed{E}_T$ event comes from the $H_5^{\pm\pm}H_3^\mp$ production.  Thus, the peak at around 150 GeV in the $M_{jj}$ distribution gives the mass of $H_3^\pm$.  Secondly, the $\ell^\pm\ell^\pm \ell^\mp jj \slashed{E}_T$ event comes from the $H_5^{\pm}H_3^\mp$ and $H_5^{\pm}H_3^0$ production processes. 
These two production cross sections are almost the same as shown in Fig.~\ref{cross}. 
Nevertheless, the decay branching fractions of $H_3^\pm\to cs$ and $H_3^0\to b\bar{b}$ are 
about 30\% and 90\%, respectively. 
Thus, the $\ell^\pm\ell^\pm \ell^\mp jj \slashed{E}_T$ event mainly comes from the $H_5^{\pm}H_3^0$ production.  Therefore, one can conclude that the peak at around 150 GeV in the $M_{jj}$ distribution is the mass of $H_3^0$.

\begin{table}[t]
\begin{center}
{\renewcommand\arraystretch{1.2}
\begin{tabular}{|c||c|c|c|c||c|c|c|c|}\hline
&\multicolumn{4}{c||}{$\ell^\pm\ell^\pm jj E_{T}\hspace{-4mm}/\hspace{4mm}$}&\multicolumn{4}{c|}
{$\ell^\pm\ell^\pm\ell^\mp jj E_{T}\hspace{-4mm}/\hspace{4mm}$} \\\hline
Cuts&$H_5^{\pm\pm}jj$&$H_5^{\pm\pm}H_3^\mp$&$W^\pm W^\pm jj$&$\mathcal{S}$&
$H_5^{\pm}jj$&$H_5^{\pm}H_3^{\mp,0}$&$W^\pm Z jj$&$\mathcal{S}$ \\\hline\hline
Basic&3.71~(8.72)&0.72~(1.63)&3.48~(8.13) &15.8~(24.1)&0.61~(1.60)&0.53~(1.21)&45.9~(1.39$\times 10^2$)&1.66~(2.36)\\\hline
$M_T$&3.65~(8.57) &0.71~(1.60)& 1.02~(2.20) & 18.8~(28.9) & 0.61~(1.60) &0.53~(1.21)& 1.16~(3.42) & 7.52~(11.3) \\\hline
\end{tabular}}
\caption{Signal and background cross sections in units of fb after each kinematic cut, along with the significance based on an integrated luminosity of 100 fb$^{-1}$. 
The numbers without (with) parentheses correspond to the case with a CM energy of $8$ TeV (14 TeV). }
\label{sig_and_bg2}
\end{center}
\end{table}

\section{Higgs to $\gamma\gamma$ and $Z\gamma$ decays \label{sec:diphoton}}

In this section, we discuss the decays of the SM-like Higgs boson $h$ to diphotons and the photon plus $Z$ boson, both of which are loop-mediated processes in the SM. 
The Higgs to diphoton decay is one of the most important modes in the Higgs boson search at the LHC. 
According to the current data, the signal strength, defined by (the observed cross section)/(the expected cross section in the SM), 
of the diphoton mode is $1.6\pm0.4$ at the CMS~\cite{Higgs_CMS} and $1.8\pm0.5$ at the ATLAS~\cite{Higgs_ATLAS}. 
It is consistent with the SM prediction at the $2\sigma$ level.  If the observed deviation persists, it may hint at contributions from new charged particles that can couple to the SM Higgs boson.  
The Higgs decay into the photon and $Z$ boson is also important to determining the structure of the Higgs sector~\cite{Chiang-Yagyu}.  
The decay rate of this mode is closely related to that of the Higgs to diphoton mode in the sense that particles contributing to the latter generally also contribute to the former.  Yet the deviations do not follow the same pattern in general \cite{Gainer,Carena_Low_Wagner,Chiang-Yagyu}.  
In the GM model, the doubly-charged Higgs boson $H_5^{\pm\pm}$ as well as the singly-charged Higgs bosons $H_5^\pm$ and $H_3^\pm$ can contribute to these processes in addition to the $W$ boson and the top quark at the one-loop level. 
The decay rates of these processes are calculated as
\begin{align}
\Gamma(h\to \gamma\gamma)
=&\frac{G_F\alpha_{\text{em}}^2m_h^3}{128\sqrt{2}\pi^3 }
\Big|\sum_{S}Q_S^2\lambda_{hSS^*}I_{0}(m_S)
+\frac{c_\alpha}{c_H}\sum_{f}Q_f^2N_c^fI_{1/2}+(c_\alpha c_H+\frac{2\sqrt{6}}{3}s_\alpha s_H)I_1\Big|^2 ,\label{hgamgam}\\
\Gamma(h\to Z\gamma)
=&\frac{\sqrt{2}G_F\alpha_{\text{em}}^2m_h^3}{128\pi^3}\left(1-\frac{m_Z^2}{m_h^2}\right)^3\notag\\
& \times \Bigg| \sum_{S}Q_S\lambda_{hSS^*}g_{ZSS^*}J_0(m_S)
+\frac{c_\alpha}{c_H}\sum_f Q_fN_f^cJ_{1/2}
+(c_\alpha c_H+\frac{2\sqrt{6}}{3}s_\alpha s_H)J_{1}\Bigg|^2, \label{hgamZ}
\end{align}
where $N_c^f=3$ (1) for $f=q$ ($\ell$), and the loop functions $I_{0,1/2,1}$ for $h\to\gamma\gamma$ and $J_{0,1/2,1}$ for $h\to Z\gamma$ are given in Appendix~C.  The summation over $S$ includes $H_5^{++}$, $H_5^+$, and $H_3^+$.
In Eqs.~(\ref{hgamgam}) and (\ref{hgamZ}), the couplings between $h$ and the charged Higgs bosons $\lambda_{hSS^*}$ are given by
\begin{align}
\lambda_{hH_5^{++}H_5^{--}}&=
\frac{2}{v^2}\Bigg\{c_Hc_\alpha(3 m_{H_3}^2-2 M_1^2)+\sqrt{\frac{2}{3}}\frac{s_\alpha}{s_H}\Big[2m_{H_5}^2+m_h^2-M_2^2
+3c_H^2(M_1^2-2m_{H_3}^2)\Big]\Bigg\},\\
\lambda_{hH_5^{+}H_5^{-}}&=-\lambda_{hH_5^{++}H_5^{--}}\\
\lambda_{hH_3^{+}H_3^{-}}&=-\frac{1}{v^2}\left[\frac{c_\alpha}{c_H} \left(2c_H^2 m_{H_3}^2+s_H^2m_h^2\right)
+\frac{2\sqrt{6}}{3}\frac{s_\alpha}{s_H} \left(2s_H^2m_{H_3}^2+c_H^2m_h^2-M_1^2\right)\right], \label{lam_hss}
\end{align}
and those between the $Z$ boson and the charged Higgs bosons $g_{ZSS^*}$ are given by 
\begin{align}
g_{ZH_5^{++}H_5^{--}}=\frac{g}{c_W}(1-2s_W^2),\quad g_{ZH_5^{+}H_5^{-}}=g_{ZH_3^{+}H_3^{-}}=-\frac{1}{2}g_{ZH_5^{++}H_5^{--}}. 
\end{align}
To illustrate how the event rates of $h\to \gamma\gamma$ and $h\to Z\gamma$ deviate from the SM predictions, we define the following ratios:
\begin{align}
R_{\gamma\gamma} = \frac{\sigma(gg\to h)_{\text{GM}}\times BR(h\to \gamma\gamma)_{\text{GM}}}{\sigma(gg\to h)_{\text{SM}}\times BR(h\to \gamma\gamma)_{\text{SM}}},\quad 
R_{Z\gamma} = \frac{\sigma(gg\to h)_{\text{GM}}\times BR(h\to Z\gamma)_{\text{GM}}}{\sigma(gg\to h)_{\text{SM}}\times BR(h\to Z\gamma)_{\text{SM}}}, 
\end{align}
where $\sigma(gg\to h)_{\text{SM}}$ [$\sigma(gg\to h)_{\text{GM}}$]
is the gluon fusion production cross section in the SM (GM model), and  
$BR(h\to X)_{\text{SM}}$ [$BR(h\to X)_{\text{GM}}$] is the branching fraction of the $h\to X$ decay mode in the SM (GM model) with $X=\gamma\gamma$ or $Z\gamma$.
In fact, the ratio in the production cross sections, $\sigma(gg\to h)_{\text{GM}}/\sigma(gg\to h)_{\text{SM}}$, can be replaced by $c_\alpha^2/c_H^2$. 

In the numerical calculation of $R_{\gamma\gamma}$ and $R_{Z\gamma}$, 
we use the parameterization given in Eq.~(\ref{mass_set}). 
The mass of the singlet Higgs boson $m_{H_1}$ does not directly affect the decay rates of $h\to \gamma\gamma$ and $h\to Z\gamma$.  Nevertheless, it affects the parameter space as constrained by the vacuum stability and unitarity conditions. 
In this parameterization, the couplings given in Eq.~(\ref{lam_hss}) can be rewritten as
\begin{align}
\lambda_{hH_5^{++}H_5^{--}}=\frac{2}{v^2}(m_{H_5}^2-M^2),\quad 
\lambda_{hH_3^{+}H_3^{-}} = -\frac{1}{c_Hv^2}(2c_H^2m_{H_3}^2+s_H^2m_h^2).
\end{align}

\begin{figure}[t]
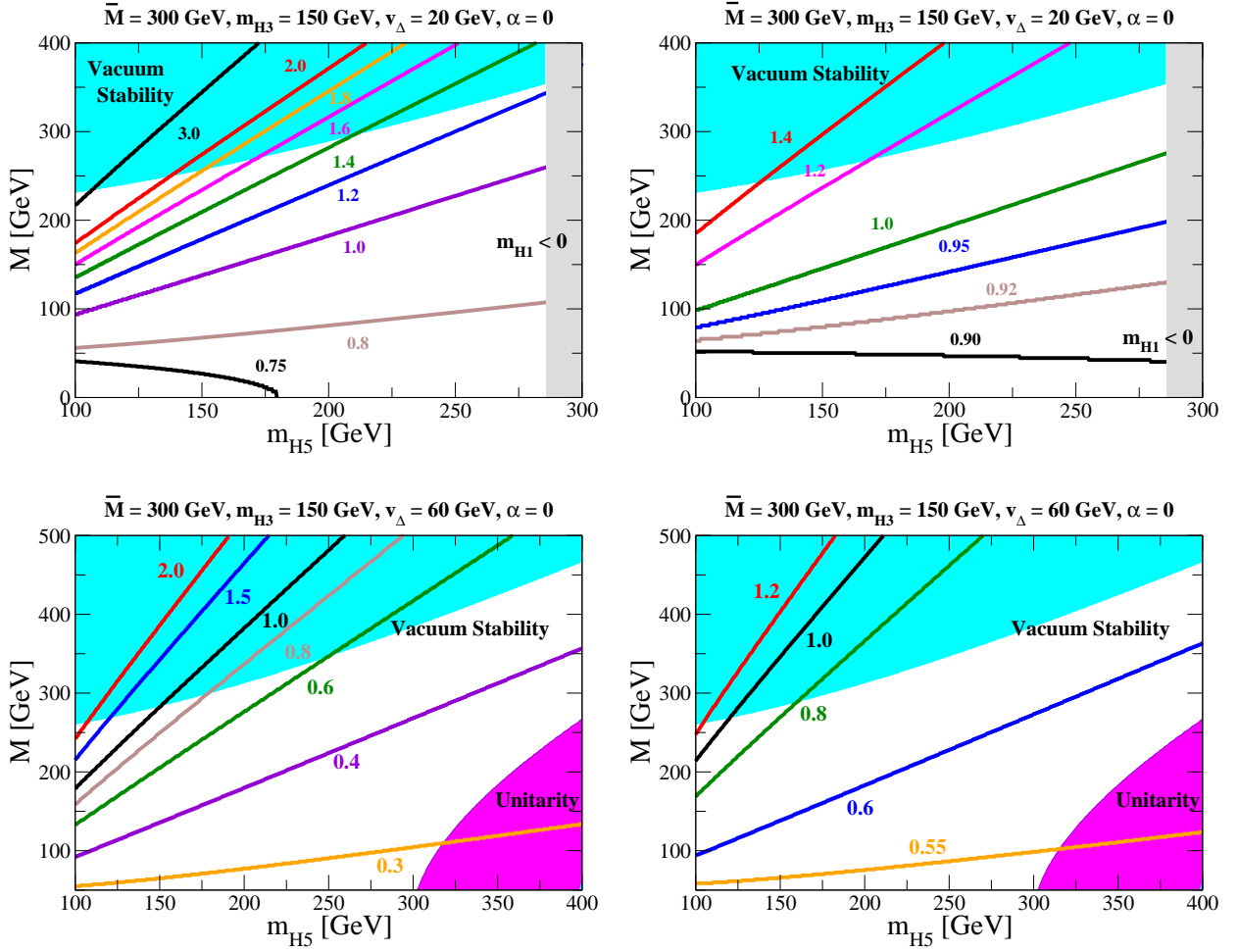

\begin{center}
\includegraphics[width=80mm]{Rgg_150_vt20_v2.eps}\hspace{3mm}
\includegraphics[width=80mm]{RZg_150_vt20_v2.eps}\\\vspace{5mm}
\includegraphics[width=80mm]{Rgg_150_vt60.eps}\hspace{3mm}
\includegraphics[width=80mm]{RZg_150_vt60.eps}
\caption{
Contour plots of $R_{\gamma\gamma}$ (left) and $R_{Z\gamma}$ (right) on the $M$-$m_{H_5}$ plane in the case of $m_{H_3}=150$ GeV, $\bar{M}=300$ GeV and $\alpha=0$.
In the upper (lower) two plots, $v_\Delta$ is taken to be 20 GeV (60 GeV).
The blue, pink and gray shaded regions are respectively 
excluded by the vacuum stability bound, unitarity bound and by having a negative mass for $H_1$. 
}
\label{R1}
\end{center}
\end{figure}

Fig.~\ref{R1} shows the contours of $R_{\gamma\gamma}$ (left plot) and $R_{Z\gamma}$ (right plot) on the $M$-$m_{H_5}$ plane.  
Here we take $m_{H_3}=150$ GeV, $\bar{M} = 300$ GeV and $\alpha = 0$. 
The triplet VEV $v_\Delta$ is taken to be 20 GeV in the upper two plots and 60 GeV in the lower two plots. 
The blue, pink and gray shaded regions are excluded by the vacuum stability bound, unitarity bound and by having a negative value for $m_{H_1}$, respectively.
For fixed values of $m_{H_5}$, both $R_{\gamma\gamma}$ and $R_{Z\gamma}$ increase with $M$.  
In other words, there is a correlation between the two ratios in this model. 
For the case with a larger $v_\Delta$, $R_{\gamma\gamma}$ and $R_{Z\gamma}$ tend to have smaller values because the $hW^+W^-$ coupling gets smaller.  
Using $m_{H_5}=150$ GeV as an example, the maximally allowed values of $R_{\gamma\gamma}$ and $R_{Z\gamma}$ are about $1.8$ ($1.0$) and $1.2$ ($0.8$) in the case of $v_\Delta=20$ GeV (60 GeV), respectively.

\begin{figure}[t]
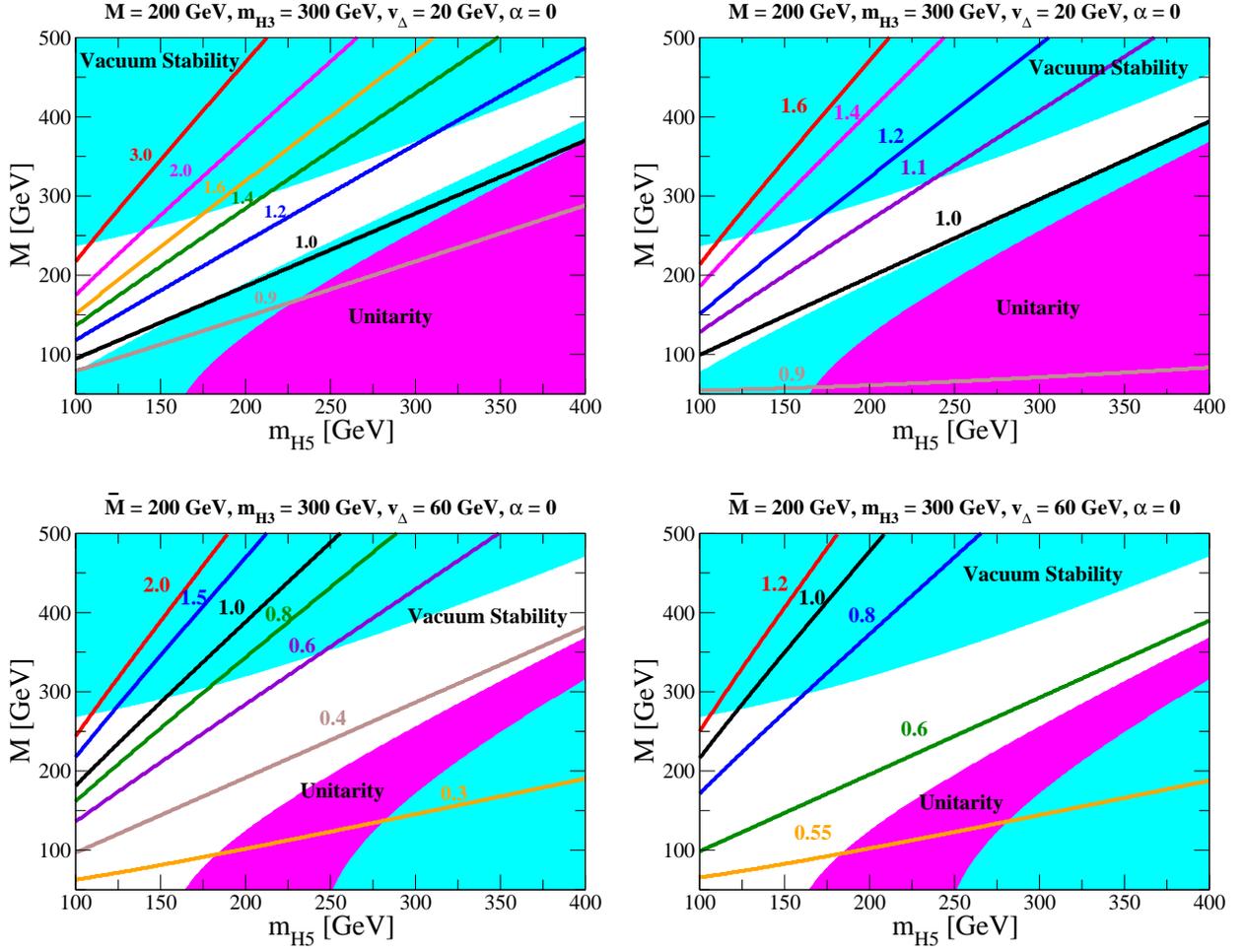

\begin{center}
\includegraphics[width=80mm]{Rgg_300_vt20.eps}\hspace{3mm}
\includegraphics[width=80mm]{RZg_300_vt20.eps}\\\vspace{5mm}
\includegraphics[width=80mm]{Rgg_300_vt60.eps}\hspace{3mm}
\includegraphics[width=80mm]{RZg_300_vt60.eps}
\caption{
Contour plots of $R_{\gamma\gamma}$ (left) and $R_{Z\gamma}$ (right) on the $M$-$m_{H_5}$ plane in the case of $m_{H_3}=300$ GeV, $\bar{M}=200$ GeV and $\alpha=0$.
In the upper (lower) two plots, $v_\Delta$ is taken to be 20 GeV (60 GeV).
The blue and pink shaded regions are respectively excluded by the vacuum stability bound and unitarity bound. }
\label{R2}
\end{center}
\end{figure}

Fig.~\ref{R2} also shows the corresponding contour plots for the case of $m_{H_3}=300$ GeV, $\bar{M}=200$ GeV and $\alpha = 0$. 
The parameter space allowed by the unitarity and the vacuum stability constraints is much smaller than the previous case. 
Again, using $m_{H_5}=150$ GeV, the maximally allowed values of $R_{\gamma\gamma}$ and $R_{Z\gamma}$ are almost the same as the case of $m_{H_3}=150$ GeV, 
but the minimum values of both $R_{\gamma\gamma}$ and $R_{Z\gamma}$ are around $1.0$ in the case of $v_\Delta = 20$ GeV.

\section{Conclusions \label{sec:summary}}

We have discussed how to test the custodial symmetry in the Higgs sector of the GM model at the LHC.  This can be done by experimentally verifying three characteristic features. 
First, there are several Higgs bosons in addition to the SM-like Higgs boson $h$; namely, 
a pair of doubly-charged Higgs bosons, two pairs of singly-charged Higgs bosons, 
a CP-odd Higgs boson and three CP-even Higgs bosons.  These Higgs bosons can be classified into 
the 5-plet Higgs bosons $(H_5^{\pm\pm}, H_5^\pm,H_5^0)$, the 3-plet Higgs bosons $(H_3^{\pm}, H_3^0)$, and the singlet Higgs boson $H_1^0$ under the custodial $SU(2)_V$ symmetry. 
The Higgs bosons belonging to the same $SU(2)_V$ multiplet have the same mass, subject to small electromagnetic corrections at the order of a few hundred MeV. 
Secondly, the 5-plet and the singlet Higgs bosons can couple to weak gauge boson pairs, but not fermion pairs via the usual (not neutrino) Yukawa interaction at the tree level. 
On the other hand, the 3-plet Higgs bosons can couple to fermion pairs, but not weak
gauge boson pairs.  As discussed in the main text, such a feature leads to specific final states for detecting these Higgs bosons and measuring their masses.
Thirdly, the VEV of the isospin triplet Higgs fields can be taken to be of 
order 10 GeV while keeping $\rho =1$ at the tree level.  This is not possible in models with triplet fields in general. 

The decay properties of the triplet-like Higgs bosons have been discussed in details. 
They depend on the mass splitting $\Delta m$, defined by $m_{H_3}-m_{H_5}$, and the triplet VEV $v_\Delta$.  We find that the parameter space in the $v_\Delta$-$\Delta m$ plane can be divided into four regions, among which the main decay modes of the triplet-like Higgs bosons are quite distinct. 

We have discussed the collider phenomenology of the GM model at the LHC in Region~II where 
the 5-plet Higgs bosons mainly decay to weak gauge boson pairs, 
whereas the main decay modes of $H_3^\pm$ are $\tau\nu$ and $cs$ and that of $H_3^0$ is $b\bar{b}$ when the mass of the 3-plet Higgs bosons is less than $m_t$. 
We focus on the VBF, the vector boson associated and the mDY production processes in order to 
verify the custodial symmetric nature of the model. 
We find that $H_5^{\pm\pm}$ and $H_5^\pm$ can be detected at more than 5$\sigma$ level
by using the forward jet tagging for the VBF process and the transverse mass cut on the 
charged leptons and missing transverse energy system if the center-of-mass energy and the luminosity are 8 TeV and 100 fb$^{-1}$, respectively. 
The significance of the $H_5^0$ Higgs boson can be reached at 3$\sigma$ level by further imposing the $b$-jet veto. 

We also find that the 3-plet Higgs bosons can be detected via the mDY production process. 
After the $M_T$ cut, 
the masses of $H_3^\pm$ and $H_3^0$ can be measured from the peak in the invariant mass distribution of the dijet system.  Therefore, the respective mass degeneracy in the 5-plet Higgs bosons and the 3-plet Higgs bosons can be tested. 

We have also investigated the $h\to \gamma\gamma$ and $h\to Z\gamma$ processes in the GM model.  
In this model, the $H_5^{\pm\pm}$, $H_5^\pm$ and $H_3^\pm$ bosons can contribute to these processes in addition to the SM top quark and the $W$ boson at one-loop level. 
We find that in the parameter space consistent with the unitarity and the vacuum stability, 
the maximally allowed value of $R_{\gamma\gamma}$ is around $1.8$ ($1.0$) for the parameter choice of $m_{H_3}=150$ GeV, $m_{H_5}=150$ GeV and $v_\Delta=20$ GeV (60 GeV). 
Deviations in the rates of $h\to Z\gamma$ and $h\to \gamma\gamma$ processes from the SM predictions can be used to distinguish models with various extended Higgs sectors. 
In the GM model, the maximally allowed value of $R_{Z\gamma}$ is around $1.2$ ($0.8$) for $m_{H_3}=150$ GeV, $m_{H_5}=150$ GeV and $v_\Delta=20$ GeV (60 GeV). 
For the cases of larger $m_{H_3}$ ({\it e.g.}, $m_{H_3}=300$ GeV), the maximally allowed values of $R_{\gamma\gamma}$ and $R_{Z\gamma}$ are not so different from the case of $m_{H_3}=150$ GeV. 
But, the minimum values of both $R_{\gamma\gamma}$ and $R_{Z\gamma}$ are about $1.0$ when $m_{H_5}=150$ GeV and $v_\Delta$=20 GeV.

\section*{Acknowledgments}
The authors would like to thank Takaaki Nomura for useful technical help.  This research was supported in part by the National Science Council of R.O.C. under Grants Nos. NSC-100-2628-M-008-003-MY4 and NSC-101-2811-M-008-014.

%
%

\begin{appendix}
\section{Relationships among different representations of the Higgs fields}

The Higgs fields expressed in Eqs.~(\ref{eq:Higgs_matrices}) and (\ref{22mat}) are related as follows: 
\begin{align}
\text{tr}(\Phi^\dagger\Phi)&=2\phi^\dagger\phi,\\
\text{tr}(\Delta^\dagger\Delta)&=2\text{tr}(\chi^\dagger\chi+\xi^\dagger\xi),\\
\text{tr}[(\Delta^\dagger\Delta)^2]&=6[\text{tr}(\chi^\dagger\chi)]^2-4\text{tr}[(\chi^\dagger\chi)^2]+2\text{tr}(\xi^4)
+4\text{tr}(\chi\xi)\text{tr}(\xi\chi^\dagger),\\
\text{tr}\left(\Phi^\dagger\frac{\tau^a}{2}\Phi\frac{\tau^b}{2}\right)\text{tr}(\Delta^\dagger t^a\Delta t^b)
&=\sqrt{2}[(\phi^\dagger \chi)(\xi\tilde{\phi})+\text{h.c.}]+2(\phi^\dagger \chi)(\chi^\dagger\phi)
-(\phi^\dagger\phi) \text{tr}(\chi^\dagger\chi),\\
\text{tr}\left(\Phi^\dagger\frac{\tau^a}{2}\Phi\frac{\tau^b}{2}\right)(P^\dagger \Delta P)^{ab}&=
-\frac{1}{\sqrt{2}}\phi^\dagger \xi\phi +\frac{1}{2}(\phi^\dagger \chi \tilde{\phi} +\text{h.c.}),\\
\text{tr}(\Delta^\dagger t^a\Delta t^b)(P^\dagger \Delta P)^{ab}&=6\sqrt{2}\text{tr}(\chi^\dagger  \chi\xi). 
\end{align}


\section{Coupling constants between the triplet-like Higgs bosons and the weak gauge bosons}
The Gauge-Gauge-Scalar vertices and the corresponding coefficients are listed in Table~\ref{GGS}. 
The Gauge-Scalar-Scalar vertices are listed in Table~\ref{GSS}, where $p_1$ and $p_2$ are respectively the four-momenta of the first and second particles in the vertex column into the vertex. 

\begin{table}[t]
{\renewcommand\arraystretch{1.5}
\begin{tabular}{|c|c||c|c|}\hline
Vertex & Coefficient & Vertex & Coefficient \\\hline
$H_5^{\pm\pm}W_\mu^{\mp}W_\nu^{\mp} $ & $\frac{g^2}{2\sqrt{2}}s_Hvg_{\mu\nu}$
&$H_1^{0}Z_\mu Z_\nu $ & $-\frac{g_Z^2}{12}(3s_\alpha c_H-2\sqrt{6}c_\alpha s_H)vg_{\mu\nu}$
\\\hline
$H_5^{\pm}W_\mu^\mp Z_\nu $ & $\mp\frac{gg_Z}{2}s_Hvg_{\mu\nu}$
&$hW_\mu^+ W_\nu^- $ & $\frac{g^2}{6}(3c_\alpha c_H+2\sqrt{6}s_\alpha s_H)vg_{\mu\nu}$\\\hline
$H_5^0W_\mu^+W_\nu^- $ & $-\frac{g^2}{2\sqrt{3}}s_Hvg_{\mu\nu}$
&$hZ_\mu Z_\nu $ & $\frac{g_Z^2}{12}(3c_\alpha c_H+2\sqrt{6}s_\alpha s_H)vg_{\mu\nu}$\\\hline
$H_5^0Z_\mu Z_\nu $ & $\frac{g_Z^2}{2\sqrt{3}}s_Hvg_{\mu\nu}$
&$G^\pm W^\mp_\mu A_\nu $ & $\pm em_W g_{\mu\nu}$\\\hline
$H_1^{0}W_\mu^+ W_\nu^- $ & $-\frac{g^2}{6}(3s_\alpha c_H-2\sqrt{6}c_\alpha s_H)vg_{\mu\nu}$
&$G^\pm W^\mp_\mu Z_\nu $ & $\mp es_Wm_Zg_{\mu\nu}$\\\hline
\end{tabular}}
\caption{Gauge-Gauge-Scalar vertices and the associated coefficients.}
\label{GGS}\vspace{5mm}
{\renewcommand\arraystretch{1.5}
\begin{tabular}{|c|c||c|c|}\hline
Vertex& Coefficient &Vertex& Coefficient\\\hline
$H_5^{++}H_5^{--}A_\mu $ & $2e(p_1-p_2)_\mu$&$H_5^{\pm\pm}H_5^{\mp}W_\mu^\mp $ & $-\frac{g}{\sqrt{2}}(p_1-p_2)_\mu$\\\hline
$H_5^{+}H_5^{-}A_\mu $ & $-e(p_1-p_2)_\mu$&$H_5^{\pm}H_5^{0}W_\mu^\mp $ & $\frac{\sqrt{3}}{2}g(p_1-p_2)_\mu$\\\hline
$H_3^{+}H_3^{-}A_\mu $ & $-e(p_1-p_2)_\mu$&$H_5^{\pm\pm}H_3^{\mp}W_\mu^\mp $ & $-\frac{g}{\sqrt{2}}c_H(p_1-p_2)_\mu$\\\hline
$G^{+}G^{-}A_\mu $ & $-e(p_1-p_2)_\mu$&$H_5^{\pm}H_3^{0}W_\mu^\mp $ & $\mp i\frac{g}{2}c_H(p_1-p_2)_\mu$\\\hline
$H_5^{++}H_5^{--}Z_\mu $ & $\frac{g}{c_W}(1-2s_W^2)(p_1-p_2)_\mu$&$H_3^{\pm}H_5^{0}W_\mu^\mp $ & $-\frac{\sqrt{3}}{6}gc_H(p_1-p_2)_\mu$\\\hline
$H_5^{+}H_5^{-}Z_\mu $ & $-\frac{g}{2c_W}(1-2s_W^2)(p_1-p_2)_\mu$&$H_3^{\pm}H_3^{0}W_\mu^\mp $ & $\mp i\frac{g}{2}(p_1-p_2)_\mu$\\\hline
$H_3^{+}H_3^{-}Z_\mu $ & $-\frac{g}{2c_W}(1-2s_W^2)(p_1-p_2)_\mu$&$H_3^{\pm}H_1^{0}W_\mu^\mp $ & 
$\frac{g}{6}(2\sqrt{6}c_Hc_\alpha+3s_Hs_\alpha)(p_1-p_2)_\mu$\\\hline
$H_5^{\pm}H_3^{\mp}Z_\mu $ & $\pm\frac{g_Z}{2}c_{H}(p_1-p_2)_\mu$&$H_3^{\pm}hW_\mu^\mp $ & $\frac{g}{6}(2\sqrt{6}c_Hs_\alpha-3s_Hc_\alpha)(p_1-p_2)_\mu$\\\hline
$H_5^0H_3^0Z_\mu $ & $i\frac{g_Z}{\sqrt{3}}c_H(p_1-p_2)_\mu$&$G^\pm h W^\mp$&$\frac{g}{6}(3c_\alpha c_H+2\sqrt{6}s_\alpha s_H)$\\\hline
$H_3^0H_1^{0}Z_\mu $ & $-i\frac{g_Z}{6}(2\sqrt{6}c_Hc_\alpha+3s_Hs_\alpha)(p_1-p_2)_\mu$&$H_3^0hZ_\mu $ & $-i\frac{g_Z}{6}(2\sqrt{6}c_Hs_\alpha-3s_Hc_\alpha)(p_1-p_2)_\mu$\\\hline
\end{tabular}}
\caption{Gauge-Scalar-Scalar vertices and the associated coefficients.}
\label{GSS}
\end{table}

\section{Loop functions in the $h \to \gamma\gamma$ and $Z\gamma$ decay rates}

The loop functions appearing in the calculations of the SM-like Higgs boson decay to diphotons are given in terms of the Passarino-Veltman functions~\cite{PV} as 
\begin{align}
I_0(m)&=\frac{2v^2}{m_h^2}[1+2m^2C_0(0,0,m_h^2,m,m,m)],\\
I_{1/2}&= -4m_f^2
\left[\frac{2}{m_h^2}-\left(1-\frac{4m_f^2}{m_h^2}\right)C_0(0,0,m_h^2,m_f,m_f,m_f)\right],\\
I_1&= 2m_W^2\left[\frac{6}{m_h^2}+\frac{1}{m_W^2}+6\left(\frac{2m_W^2}{m_h^2}-1\right)C_0(0,0,m_h^2,m_W,m_W,m_W)\right]. 
\label{I_func}
\end{align}
Those for the $h\to Z\gamma$ process are given by
\begin{align}
J_0(m)=&
\frac{2}{e(m_h^2-m_Z^2)}\Big[
1+2m^2C_0(0,m_Z^2,m_h^2,m,m,m)+\frac{m_Z^2}{m_h^2-m_Z^2}(B_0(m_h^2,m,m)-B_0(m_Z^2,m,m))\Big],\\
J_{1/2}=&
-\frac{4m_f^2(\frac{1}{2}I_f-s_W^2Q_f)}{s_Wc_W(m_h^2-m_Z^2)}\Big[2+(4m_f^2-m_h^2+m_Z^2)C_0(0,m_Z^2,m_h^2,m_f,m_f,m_f)\notag\\
&+\frac{2m_Z^2}{m_h^2-m_Z^2}(B_0(m_h^2,m_f,m_f)-B_0(m_Z^2,m_f,m_f))\Big],\\
J_1=&
\frac{2m_W^2}{s_Wc_W(m_h^2-m_Z^2)}\Bigg\{\left[c_W^2\left(5+\frac{m_h^2}{2m_W^2}\right)-s_W^2\left(1+\frac{m_h^2}{2m_W^2}\right)\right]\notag\\
& \times\left[1+2m_W^2C_0(0,m_Z^2,m_h^2,m_W,m_W,m_W)+\frac{m_Z^2}{m_h^2-m_W^2}(B_0(m_h^2,m_W,m_W)-B_0(m_Z^2,m_W,m_W))\right]\notag\\
&-6c_W^2(m_h^2-m_Z^2)C_0(0,m_Z^2,m_h^2,m_W,m_W,m_W)+2s_W^2(m_h^2-m_Z^2)C_0(0,m_Z^2,m_h^2,m_W,m_W,m_W)\Bigg\}. 
\label{J_func}
\end{align}

\end{appendix}

\end{document}